\documentclass[twoside,11pt]{article}

\usepackage{arxiv}

\usepackage{amsmath,amssymb,amsfonts,amsthm}
\usepackage{graphicx}
\usepackage{booktabs}
\usepackage{multirow}
\usepackage{xcolor}
\usepackage{algorithm}
\usepackage{algorithmicx}
\usepackage{algpseudocode}
\usepackage{listings}
\usepackage{comment}
\usepackage[numbers]{natbib}
\usepackage[
  colorlinks=true,
  linkcolor=blue,
  citecolor=blue,
  urlcolor=blue
]{hyperref}
\usepackage[title]{appendix}
\usepackage{fancyhdr}
\usepackage{float}

\usepackage[noframe]{showframe}
\usepackage{framed}
\usepackage{lipsum}
\renewenvironment{shaded}{%
  \MakeFramed{\advance\hsize-\width \FrameRestore\FrameRestore}}%
 {\endMakeFramed}
\definecolor{shadecolor}{gray}{0.75}

\title{Meta-analysis and network meta-analysis of time-to-event outcomes with non-proportional hazards: a Bayesian time-varying hazard ratio approach\thanks{Presented as an Oral Contribution at ISCB-46, Basel, August 27, 2025}}

\author{
{\bf Rhiannon K.\ Owen}\\Population Data Science, \\Swansea University Medical School, \\Swansea University, UK.\\
Email: \textcolor{blue}{\texttt{r.k.owen@swansea.ac.uk}}
\and
{\bf Keith R.\ Abrams}\\Department of Statistics \& \\ 
Warwick Medical School (WMS), \\University of Warwick, UK.\\
Email: \textcolor{blue}{\texttt{Keith.Abrams@warwick.ac.uk}}
}
\begin{document}

\maketitle

% ---------- Abstract ----------
\begin{abstract}
\textbf{Background:} Often when undertaking meta-analyses of time-to-event (TTE) outcomes, especially in a Health Technology Assessment (HTA) context, a hazard ratio (HR) scale is used. However, issues arise when there is evidence of non-proportional hazards in some of the trials/studies included. Whilst a number of methods have been advocated in such situations, their use has been limited by either their complexity and/or the ease with which their results can be incorporated into an economic decision model in order to assess cost-effectiveness. 
 
An alternative approach is to assume a treatment-log(time) interaction within a Cox proportional hazards model for each trial/study, thus allowing the log HR to vary linearly with respect to log(time), and to then undertake a bivariate meta-analysis of the resulting treatment and interaction coefficients, so that an overall time-varying HR (TVHR) can be obtained with appropriate uncertainty.

 \textbf{Methods:} A TVHR approach was applied to a meta-analysis of 20 trials, involving 4,069 patients, of chemotherapy compared to Standard of Care (SoC) for advanced recurrent gastric cancer, and in which Progression-Free Survival (PFS) was an outcome with median follow-up of 1.2 years. The approach was also applied to a network meta-analysis (involving 13 treatments across 13 RCTs evaluating overall survival (OS)) in previously untreated advanced BRAF-mutated melanoma and in which there were 3,913 all-cause deaths in 6,378 participants. In both examples the approach was compared with a standard Bayesian meta-analysis or network meta-analysis under the assumption of proportional hazards. 
 
 \textbf{Results:} Five trials in the advanced gastric cancer meta-analysis displayed evidence of non-proportional hazards for PFS. A standard Bayesian random effects meta-analysis of HRs yielded a pooled HR of 0.78 (95\% CrI: 0.70 to 0.86). Using a TVHR model produced a pooled interaction effect of 0.09 (95\% CrI: -0.005 to 0.193) with resulting HRs ranging from 0.83 (95\% CrI: 0.75 to 0.91) at 0.5 years to 0.99 (95\% CrI: 0.79 to 1.23) at 3.5 years. 
 
 Three studies showed evidence of non-proportional hazards in the advanced BRAF-mutated melanoma network meta-analysis for OS. Using a TVHR model, nivolumab plus ipilimumab demonstrated consistent superiority from month 7 onwards, with a HR improving from 0.37 (95\% CrI: 0.26 to 0.51) at one year to 0.24 (95\% CrI: 0.12–0.45) at five years, retaining the highest mean rank at year 5 (mean rank: 1.18, 95\% CrI: 1 to 2). Dabrafenib and vemurafenib based regimens displayed evidence of treatment effect waning and may be potentially inferior (HR$>$1) at 5 years compared to dacarbazine.  
 
 \textbf{Conclusions:} A TVHR approach to the meta-analysis or network meta-analysis of TTE outcomes when the proportional hazards assumption appears not to hold, at least for some of the studies included, produces a simple and intuitive solution which can be readily incorporated into an economic decision model. Further extension to a fully Bayesian one-stage model and incorporating splines on the interaction effect for more flexibility are also possible. 
\end{abstract}

\bigskip
\noindent\textbf{Keywords:}
Meta-analysis; Network meta-analysis; Bayesian methods; Non-proportional hazards; Time-varying hazard ratio

%%%%%%%%%%%%%%%%%%%%%%%%%%%%%%%%%%%%%%%%%%%%%%%%%%%%%%%%
% INTRO & BACKGROUND

\section{Introduction \& Background}\label{intro}

Often when conducting meta-analyses (MA) of time-to-event (TTE) results, especially in Health Technology Assessment (HTA), a hazard ratio (HR) scale is used due to the ease with which a treatment effect can be incorporated into a cost-effectiveness decision model \citep{cope2023,Koblbauer2024}. However, issues arise when there is evidence of non-proportional hazards in some of the trials/studies included. In a network meta-analysis setting, departures from proportional hazards introduce additional complexity, as the assumption of consistency between direct and indirect treatment comparisons may be violated when HRs vary over time \citep{jansen2011}.

When Individual Participant Data (IPD) are available, or have been re-created from published Kaplan-Meier curves \citep{guyot2012}, a number of methods have been advocated, including; parametric models \citep{ouwens2010,campbell2025}, flexible parametric models \citep{freeman2017}, piecewise exponential models \citep{freeman2024}, fractional polynomial models \citep{jansen2011}, flexible M-splines \citep{phillippo2025}, and Restricted Mean Survival Time (RMST) models \citep{deJong2020}. However, their use has been limited by either their complexity and/or the ease with which their results can be incorporated into an economic decision model in order to assess cost-effectiveness. 

In a HTA setting, a HR is usually applied to a target baseline population (with or without treatment waning) \citep{cope2023,Koblbauer2024}. Whilst some of the existing methods enable a HR to be estimated others do not. Treatment effect waning scenarios are also more easily incorporated and interpreted on a HR scale, as it is a relative measure \citep{trigg2024treatment}. 

In this paper we present a simple Time-Varying Hazard Ratio (TVHR) approach - which assumes that the HR in each study in a meta-analysis can change linearly with respect to log(time) and the resulting coefficients from each study are then synthesised using a bivariate meta-analysis to yield an overall relationship between the log HR and log(time) \citep{therneau2024}. A key feature of the proposed framework is that the standard proportional hazards meta-analysis arises as a special case when the coefficient on log(time) is zero.  

The rest of the paper is organised as follows; Section~\ref{examples} introduces two motivating examples - one a pairwise meta-analysis of treatments for advanced gastric cancer, and the second a network meta-analysis of 13 treatments for advanced BRAF-mutated Melanoma - in both cases there was evidence of non-proportional hazards in some of the studies included. Section~\ref{methods} describes the methods which can be used to undertake a pairwise or network meta-analysis when either proportional hazards is assumed to hold or when the assumption is relaxed and a TVHR approach adopted, whilst Section~\ref{results} presents the results of applying a TVHR approach to the two motivating examples. Finally, Sections~\ref{discussion} and \ref{conclusion} discuss some of the advantages and limitations of a TVHR approach and present conclusions. 

%%%%%%%%%%%%%%%%%%%%%%%%%%%%%%%%%%%%%%%%%%%%%%%%%%%%%%%%
% MOTIVATING EXAMPLES

\section{Motivating Examples}\label{examples}

% PAIRWISE - GASTRIC CANCER
%
\subsection{Pairwise meta-analysis - Advanced Gastric Cancer}
\begin{figure}[htbp]
  \centering
  \includegraphics[width=0.8\textwidth]{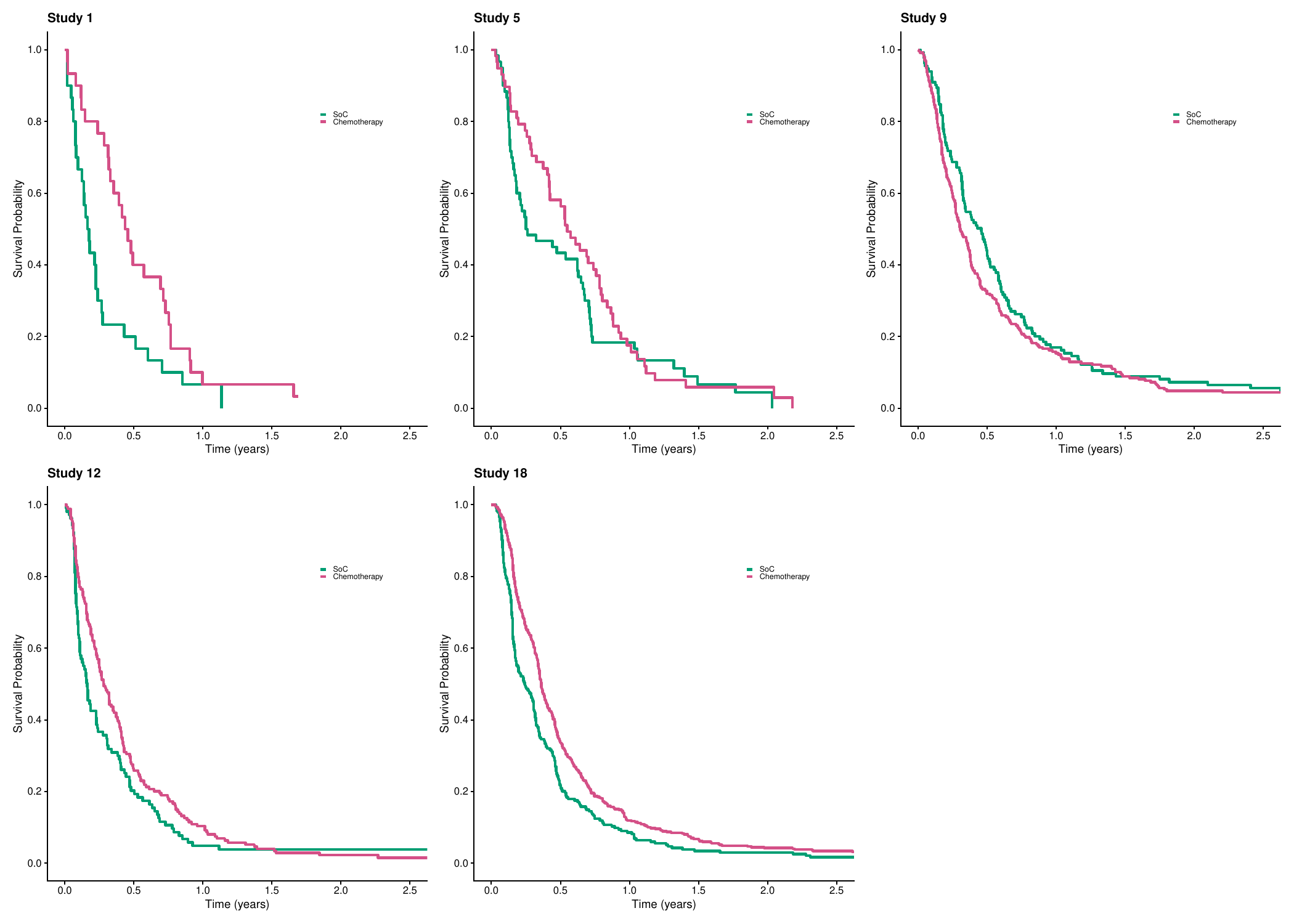}
  \caption{Gastric cancer meta-analysis - Kaplan-Meier curves for studies with non-proportional hazards.}
  \label{fig:gastricnonPHKM}
\end{figure}

An IPD meta-analysis of 20 trials, ranging in size from 58 patients to 704 patients, involving 4,069 patients, comparing chemotherapy to Standard of Care (SoC) for advanced recurrent gastric cancer was undertaken by the Global Advanced/Adjuvant Stomach Tumour Research International Collaboration (GASTRIC) Group \citep{gastric2013} \& \citep{rotolo2018}. The primary outcome was Progression-Free Survival (PFS) with a median follow-up of 1.2 years and up to a maximum of 7.2 years during which 3,820 events were observed. Of the 20 trials in the meta-analysis 5 (Study 1, P=0.04; Study 5, P=0.09; Study 9, P=0.04; Study 12, P=0.02 \& Study 18, P=0.0004) displayed evidence of non-proportional hazards for PFS as can also be seen from the Kaplain-Meier and Schoenfeld residual plots in Figures~\ref{fig:gastricnonPHKM}~\&~\ref{fig:gastricnonPH}. However, it should be noted that formal testing for departures from the assumption of proportional hazards has been shown to have low statistical power  \citep{austin2018statistical}.  

% NMA - MELANOMA 
%
\subsection{Network meta-analysis - Advanced BRAF-mutated Melanoma}
\begin{figure}[htbp]
  \centering
  \includegraphics[width=0.8\textwidth]{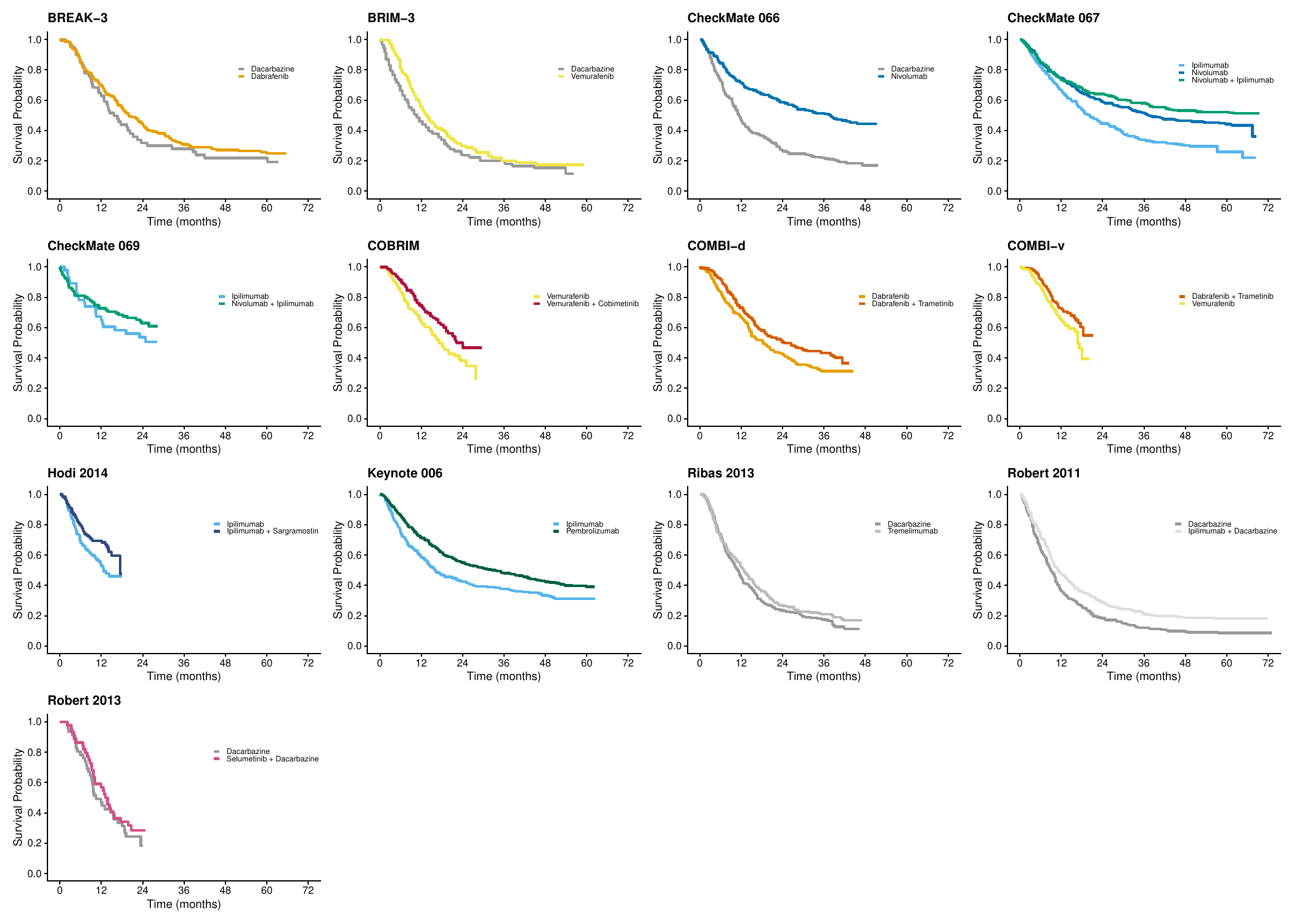}
  \caption{Advanced BRAF-mutated melanoma NMA - Kaplan-Meier Plots.}
  \label{fig:KM_NMA}
\end{figure}

A network meta-analysis in previously untreated advanced BRAF-mutated melanoma as discussed in \citep{zoratti2019} with full IPD extracted and reported in  \citep{freeman2022} was used as a motivating example. The primary outcome was overall survival with median follow-up ranging from 1 year to 5.3 years. The network included 13 RCTs and 13 treatments including dabrafenib (DB), dacarbazine (DTIC), ipilimumab (IPI), tremelimumab (TRL), nivolumab (NIV), pembrolizumab (PEM), and vemurafenib (VM), as single treatments, and dabrafenib + trametinib (DB + TR), ipilimumab + dacarbazine (IPI + DTIC),  ipilimumab + sargramostin (IPI + SRG), nivolumab + ipilimumab (NIV + IPI), selumetinib + dacarbazine (SEL + DTIC) and vemurafenib + cobimetinib (VM + COB), as combination treatments (Figure~\ref{fig:melanomNMA}). There were 12 two-arm trials, and 1 three-arm trial included in the network with the majority of direct treatment comparisons informed by single studies. In total, 6,378 participants were included in the network, with 3,913 mortality events. The sample size of included studies ranged from 45 to 556 participants. Figure \ref{fig:KM_NMA} illustrates the Kaplan-Meier plots for all 13 studies. Inspection of Schoenfeld residual plots suggested possible time-varying effects across several studies (Figure \ref{fig:zph_NMA}). However, evidence from global tests of proportional hazards indicated statistically significant deviations at the 5\% level in three studies (BRIM-3, CheckMate-067, and Keynote 006).

%%%%%%%%%%%%%%%%%%%%%%%%%%%%%%%%%%%%%%%%%%%%%%%%%%%%%%%%
% METHODS 
%%%%%%%%%%%%%%%%%%%%%%%%%%%%%%%%%%%%%%%%%%%%%%%%%%%%%%%%

\section{Methods}\label{methods}

%%%%%%%%%%%%%%%%%%%%%%%%%%%%%%%%%%%%%%%%%%%%%%%%%%%%%%%%
% SURVIVAL MODELS 

\subsection{Proportional Hazards Cox Models \& Time-varying Hazard Ratio Models}
The standard Cox Proportional hazards regression model is given by (\ref{eq:cox}) in which $x_i$ represents a binary treatment indicator and $\beta$ the log-hazard ratio,
\begin{equation} 
\label{eq:cox}
\lambda(t) = \lambda_0(t) \exp(\beta x_i)
\end{equation}
and $\lambda_0(t)$ is an unspecified baseline hazard function with $\beta$ estimated via a partial likelihood \citep{collett2023modelling}. 
The standard Cox PH model in (\ref{eq:cox}) can be extended to allow for a time-varying (log-)hazard ratio by including an interaction term between the treatment indicator and log(time) \cite{therneau2024}, 
\begin{equation} 
\label{eq:coxtvhr}
\lambda(t) = \lambda_0(t) \exp(\beta_1 x_i + \beta_2 x_i \log(t))
\end{equation}
The hazard ratio at time $t$ is then given by $\exp(\hat{\beta_1}+\hat{\beta_2} \log(t))$. 
Models such as (\ref{eq:coxtvhr}) can be easily fit in {\tt R} \cite{R} using the following code \cite{therneau2024}; 
\begin{shaded}
\begin{verbatim}
> coxph(Surv(time, event) ~ x + tt(x), data=dat, 
             tt = function(x, t, ...) x * log(t)) 
\end{verbatim}   
\end{shaded}
where the variables {\tt time}, {\tt event} and {\tt x} present the time to an event, event indicator and binary covariate respectively in the data frame, {\tt dat}. 

%%%%%%%%%%%%%%%%%%%%%%%%%%%%%%%%%%%%%%%%%%%%%%%%%%%%%%%%
% PAIRWISE MA 

\subsection{Pairwise Meta-Analysis Model}
\subsubsection{Proportional Hazards Cox Models}
For study $i = 1,\dots,n_s$ with two arms, let
$y_i$ denote the estimated log hazard ratio for the treatment
arm compared to control, obtained from Cox proportional hazards models such as (\ref{eq:cox}). We assume
\begin{equation}
\label{eq:pma1}
y_i \sim \mathrm{N}\!\left(\delta_i,\, S_i\right),
\qquad i=1,\dots,n_s ,
\end{equation}
where $S_i$ is the estimated within-study variance which is assumed to be known. The true relative treatment effects, $\delta_i$, are modelled using a standard
pairwise meta-analysis random-effects formulation;
\begin{equation}
\label{eq:pma2}
\delta_i
\sim
\mathrm{N}\!\left(
d\; , \;
\tau^2
\right),
\end{equation}

Adopting a Bayesian formulation, the overall pooled log hazard ratio, $d$, is often assumed to have a relatively vague prior distribution such that, $d \sim \mathrm{N}(0,1000)$, whilst the between-study standard deviation, $\tau$, can be assumed to have a $\mathrm{Unif}(0,1)$ prior distribution. Such a $\mathrm{Unif}(0,1)$ prior distribution implies that the HRs of two random studies could differ by a factor of 2.97 and the range of minimum to maximum HRs could differ by a factor of 50 \citep{spiegelhalter2004bayesian}. 
Other prior distributions for $\tau$ could include informative prior distributions based on meta-research \citep{rover2021prior}. Estimation of $d$ and $\tau$ is most usually undertaken via Markov Chain Monte Carlo (MCMC) \citep{sutton2001bayesian}.   

\subsubsection{Time-varying Hazard Ratio Models}
%%%%%%%%%%%%%%%%%%%%%%%%%%%%%%%%%%%%%%%%%%%%%%%%%%%%%%%%%%%%
% Bivariate Pairwise Random-Effects Meta-Analysis
%%%%%%%%%%%%%%%%%%%%%%%%%%%%%%%%%%%%%%%%%%%%%%%%%%%%%%%%%%%%

For each study $i = 1,\dots,n_s$, let $\mathbf{y}_i$
denote the bivariate vector of estimated coefficients from a model such as (\ref{eq:coxtvhr}) with  
$\mathbf{y}_i = (y_{i1},y_{i2})$ where $y_{i1}$ is the log hazard ratio for treatment, and $y_{i2}$ is the 
treatment-log(time) interaction. Further, let the within-study covariance matrix be $\mathbf{S}_i$ and which is assumed to be known. Assuming multivariate Normality (MVN) on the log hazard ratio scale the likelihood is thus 
\begin{equation}
\label{eq:pma3}
    \mathbf{y}_i \sim \textrm{MVN}\!\left(
    \boldsymbol{\delta}_{i},\;
    \mathbf{S}_i
    \right),
    \qquad i=1,\dots,n_s .
\end{equation}
where $\boldsymbol{\delta}_{i} = (\delta_{i1},\delta_{i2})$ represents the vector of true relative effects for treatment log hazard ratio and the treatment-log(time) interaction.
%%%%%%%%%%%%%%%%%%%%%%%%%%%%%%%%%%%%%%%%%%%%%%%%%%%%%%%%%%%%
% Random-effects model
%%%%%%%%%%%%%%%%%%%%%%%%%%%%%%%%%%%%%%%%%%%%%%%%%%%%%%%%%%%%

Further $\boldsymbol{\delta}_{i}$ is assumed to follow a bivariate random-effects model:
\begin{equation}
\label{eq:pma4}
\boldsymbol{\delta}_{i}
\sim
\textrm{MVN}\!\left(
\mathbf{d},\; \boldsymbol{\Sigma} \right) 
\end{equation}
in which $d_1$ and $d_2$, the pooled treatment log hazard ratio and treatment-log(time) interaction respectively, are assumed to have independent $N(0,1000)$ vague prior distributions and 
the $2 \times 2$ between-study covariance matrix, $\boldsymbol{\Sigma}$,  is defined such that 
\begin{equation}
\label{eq:pma5}
\boldsymbol{\Sigma}
=
\begin{pmatrix}
\tau_1^2 & \rho\, \tau_1 \tau_2 \\
\rho\, \tau_1 \tau_2 & \tau_2^2
\end{pmatrix},
\end{equation}
and is estimated using a spherical parameterisation under an assumption of homogeneous between-study variances and correlation \citep{wei2013bayesian} such that
\begin{align}
\tau_m &\sim \textrm{Unif}(0,1), \qquad m=1,2, \\
\theta &\sim \textrm{Unif}(0,\pi), \qquad \rho = \cos(\theta).
\end{align}

In order to explore sensitivity to the prior distribution specification for the treatment-log(time) interaction parameter, $\tau_2$, in the between-study covariance, $\boldsymbol{\Sigma}$, a Half-Normal(0,0.5) prior distribution can also used \citep{gelman2006prior}.

%%%%%%%%%%%%%%%%%%%%%%%%%%%%%%%%%%%%%%%%%%%%%%%%%%%%%%%%
% NMA METHODS 
%%%%%%%%%%%%%%%%%%%%%%%%%%%%%%%%%%%%%%%%%%%%%%%%%%%%%%%%

\subsection{Network meta-analysis}
\subsubsection{Proportional Hazards Cox Models}
For study $i = 1,\dots,n_s$ with $n_a(i)$ arms, let $y_{i,(k-1)}$ denote the estimated log hazard ratio for arm $k=2,\dots,n_a(i)$ relative to the study-specific reference arm ($k=1$), obtained from Cox proportional hazards models. We assume

\begin{equation}
\begin{aligned}
y_{i,(k-1)} \sim \mathrm{N}\!\left(\delta_{i,t[i,k]}, \; S_{i,(k-1)}\right),
\quad k=2,\dots,n_a(i),
\end{aligned}
\end{equation}

where $S_{i,(k-1)}$ is the known within-study variance. The true relative treatment effects are modelled using a conditional formulation to account for correlations induced by multi-arm trials:

\begin{equation}
\begin{aligned}
\delta_{i,t[i,k]} \sim 
\mathrm{N}\Bigg(
d_{t[i,k]} - d_{t[i,1]} 
+ \frac{1}{k-1} \sum_{a=1}^{k-1} w_{i,a},
\; \frac{2(k-1)}{k}\tau^2
\Bigg),
\quad k=2,\dots,n_a(i),
\end{aligned}
\end{equation}

where the study-specific random effects are defined as

\begin{equation}
\begin{aligned}
w_{i,k} &= \delta_{i,t[i,k]} - \big(d_{t[i,k]} - d_{t[i,1]}\big),
\quad k=1,\dots,n_a(i), \\
\delta_{i,t[i,1]} &= 0.
\end{aligned}
\end{equation}

The basic parameters $d_t$ represent pooled log hazard ratios relative to treatment $1$, with prior distributions

\begin{equation}
\begin{aligned}
d_t &\sim \mathrm{N}(0,1000), \quad t=2,\dots,n_t, \\
d_1 &= 0.
\end{aligned}
\end{equation}

Between-study heterogeneity is captured by a common variance parameter, $\tau$, such that 

\begin{equation}
\begin{aligned}
\tau \sim \mathrm{Unif}(0,1).
\end{aligned}
\end{equation}

\subsubsection{Time-varying Hazard Ratio Models}
Let $\mathbf{y}_{i,(k-1)}$ be the estimated $m$-vector of coefficients for arm $k = 2,...,n_a(i)$ versus the study-specific reference arm, $k=1$ such that $m=1$ represents the log hazard ratio for the treatment effect, and $m=2$ the interaction of treatment and log-time, obtained from Cox proportional hazards models for study $i = 1,2,...,n_s$. Let $\mathbf{y}_{i,(k-1)}$ follow a multivariate normal distribution, with known contrast-specific within-study covariance, $\mathbf{S}_{i,(k-1)}$, obtained from Cox proportional hazards models such that:
\begin{equation}
\begin{aligned}
    \mathbf{y}_{i,(k-1)} &\sim \textrm{MVN}\!\left(\boldsymbol{\delta}_{i,t[i,k]}, \mathbf{S}_{i,(k-1)}\right), \quad k=2,\dots,n_a(i). \\[6pt]
\end{aligned}
\end{equation} 

The true coefficients for the relative treatment effect, $\boldsymbol{\delta}_{i,t[i,k]}$, for treatment $t[i,k]$, in study $i$ arm $k$ is given by

\begin{equation}
\begin{aligned}
\boldsymbol{\delta}_{i,t[i,k]} &\sim \textrm{MVN} \Bigg(
\mathbf{d}_{t[i,k]} - \mathbf{d}_{t[i,1]} 
+ \frac{1}{(k-1)} \sum_{a=1}^{(k-1)} \mathbf{w}_{i,a}, \;
\frac{2(k-1)}{k}\boldsymbol{\Sigma}
\Bigg), \quad k=2,\dots,n_a(i)
\end{aligned}
\end{equation}

The conditional covariance for arm $k$ is given by $\frac{2(k-1)}{k}\boldsymbol{\Sigma}$. For $k=2$ this reduces to $\Sigma$, which is the marginal of the first contrast. The conditional mean is calculated using the average of the previous $\textbf{w}_{i,a}$ to apply an appropriate multi-arm correction to the covariance structure, where

\begin{equation}
\begin{aligned}
    \mathbf{w}_{i,k} &= \boldsymbol{\delta}_{i,t[i,k]} - (\mathbf{d}_{t[i,k]} - \mathbf{d}_{t[i,1]}),\quad k=1,\dots,n_a(i) \\[6mm]
\end{aligned}
\end{equation}

\begin{equation}
\begin{aligned}
    \boldsymbol{\delta}_{i,t[i,1]} &= \mathbf{0}, 
\end{aligned}
\end{equation}

The $m$-vector of pooled treatment effects for treatment, $t[i,k]$, in arm $k$ of study $i$, relative to the study-specific reference arm, $t[i,1]$, is indicated by $\mathbf{d}_{t[i,k]}$, with non-informative prior distributions assigned to the basic parameters for each coefficient $m$ such that
\begin{equation}
\begin{aligned}
    d_{t,m} &\sim \textrm{N}(0, 1000), \quad \textrm{for} \quad t=2,\dots,\textrm{$n_t$}, \\[1pt]
    d_{1,m} &= 0 \quad m=1,2,
\end{aligned}
\end{equation}

The $m \times m$ between-study covariance, $\boldsymbol{\Sigma}$, is estimated using a spherical parameterisation under an assumption of homogeneous between-study variances \citep{lumley2002network} and correlations across treatment contrasts \citep{achana2014network} such that
%%%%%%%%%%%%%%%%%%%%%%%%%%%%%%%%%%%%%%%%%%%%%%%%%%%%%%%%%%%%
% --- Between-study covariance ---
%%%%%%%%%%%%%%%%%%%%%%%%%%%%%%%%%%%%%%%%%%%%%%%%%%%%%%%%%%%%
\begin{align}
\boldsymbol{\Sigma} &=
\begin{pmatrix}
\tau_1^2 & \rho \, \tau_1 \tau_2 \\
\rho \, \tau_1 \tau_2 & \tau_2^2
\end{pmatrix}, \\
&
\tau_m \sim \text{Unif}(0,1), \quad m=1,2, \\
\theta &\sim \text{Unif}(0,\pi), \quad \rho = \cos(\theta).
\end{align}

%%%%%%%%%%%%%%%%%%%%%%%%%%%%%%%%%%%%%%%%%%%%%%%%%%%%%%%%%%%%
% --- Model fitting ---
%%%%%%%%%%%%%%%%%%%%%%%%%%%%%%%%%%%%%%%%%%%%%%%%%%%%%%%%%%%%

\subsection{Model Fitting}
Models were fitted in {\tt R} \cite{R} and {\tt JAGS} \cite{plummer2003jags} using {\tt rjags} \cite{plummer2025rjags}. All models were run for 50,000 iterations of burn-in followed by 50,000 posterior samples across three chains, with a thinning factor of 5, yielding a retained sample of 10,000 draws per chain (30,000 draws across chains). The convergence of MCMC chains was assessed for basic parameters using the Gelman-Rubin potential scale reduction factor (PSRF), $\hat{R}$, calculated for each monitored parameter in three independent chains \citep{gelman1992inference}. The PSRF compares between-chain variance to within-chain variance, with values close to 1.00 indicating that the chains have converged to a common stationary distribution. Values of $\hat{R} <$ 1.01 were used as the threshold for convergence \citep{vehtari2021rank}. Effective sample size (ESS) was computed for all monitored parameters using the \texttt{effectiveSize} function in the \texttt{coda} package \citep{plummer2006coda}, with a minimum ESS of 1000 per parameter considered sufficient for reliable posterior inference with regards to means and credible intervals. Convergence of MCMC chains was visually assessed through autocorrelation and trace plots \citep{plummer2006coda}. All analyses were undertaken using an annual time scale in order to ensure adequate mixing of MCMC chains. All code for fixed and random effects models are provided at: \url{https://github.com/r-k-owen/NMA-TVHR}. 

%%%%%%%%%%%%%%%%%%%%%%%%%%%%%%%%%%%%%%%%%%%%%%%%%%%%%%%%
% RESULTS
%%%%%%%%%%%%%%%%%%%%%%%%%%%%%%%%%%%%%%%%%%%%%%%%%%%%%%%%

\section{Results}\label{results}

%%%%%%%%%%%%%%%%%%%%%%%%%%%%%%%%%%%%%%%%%%%%%%%%%%%%%%%%
% PAIRWISE RESULTS

\subsection{Pairwise meta-analysis - Advanced Gastric Cancer}
Applying a Bayesian 2-stage meta-analysis using (\ref{eq:pma1}) \& (\ref{eq:pma2}) to the advanced gastric cancer meta-analysis produces an overall posterior HR of 0.78 with 95\% CrI (0.70 to 0.86) (Table~\ref{tab:gastricHR}). These results were similar to both a frequentist 2-stage and 1-stage meta-analysis (see \ref{SM_PMA}). The very slightly wider interval obtained using a Bayesian approach being due to allowing for the uncertainty in the estimation of the between-study variance, $\tau^2$.   

\begin{figure}[htbp]
  \centering
  \includegraphics[width=0.8\textwidth]{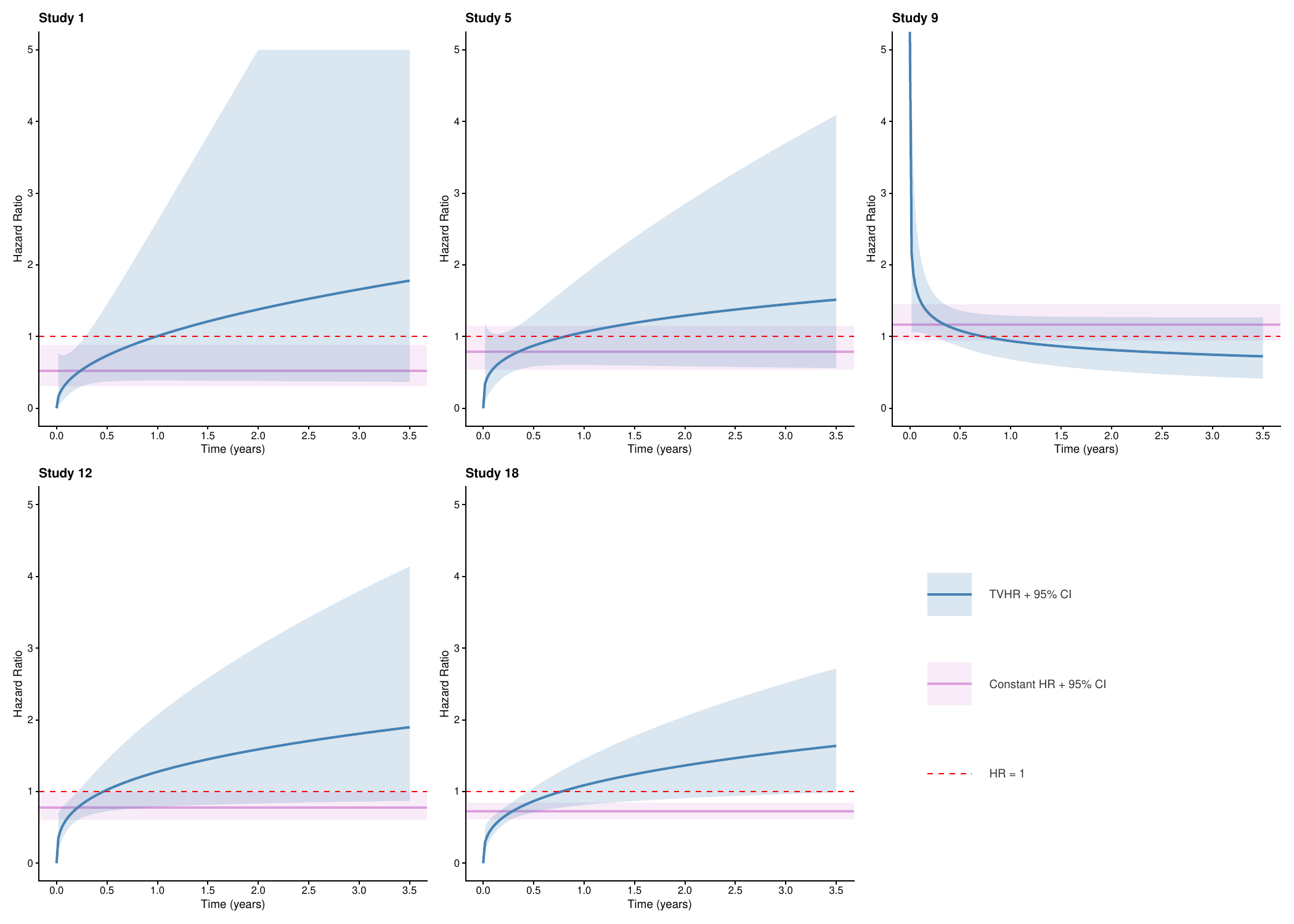}
  \caption{Gastric cancer meta-analysis - study-specific time-varying hazard ratios (TVHR) and constant hazard ratios, and associated 95\% confidence intervals (CI).}
  \label{fig:gastricstudytvhr}
\end{figure}

Figure~\ref{fig:gastricstudytvhr} displays the results of fitting a TVHR model (\ref{eq:coxtvhr}) to each of the 5 studies which display some evidence of non-proportional hazards in Figures~\ref{fig:gastricnonPHKM}~\&~\ref{fig:gastricnonPH} compared to a standard Cox model (\ref{eq:cox}) and it can be seen that there are substantial discrepancies in terms of the resulting HRs at both earlier and later time points.

\begin{figure}[htbp]
  \centering
  \includegraphics[width=0.8\textwidth]{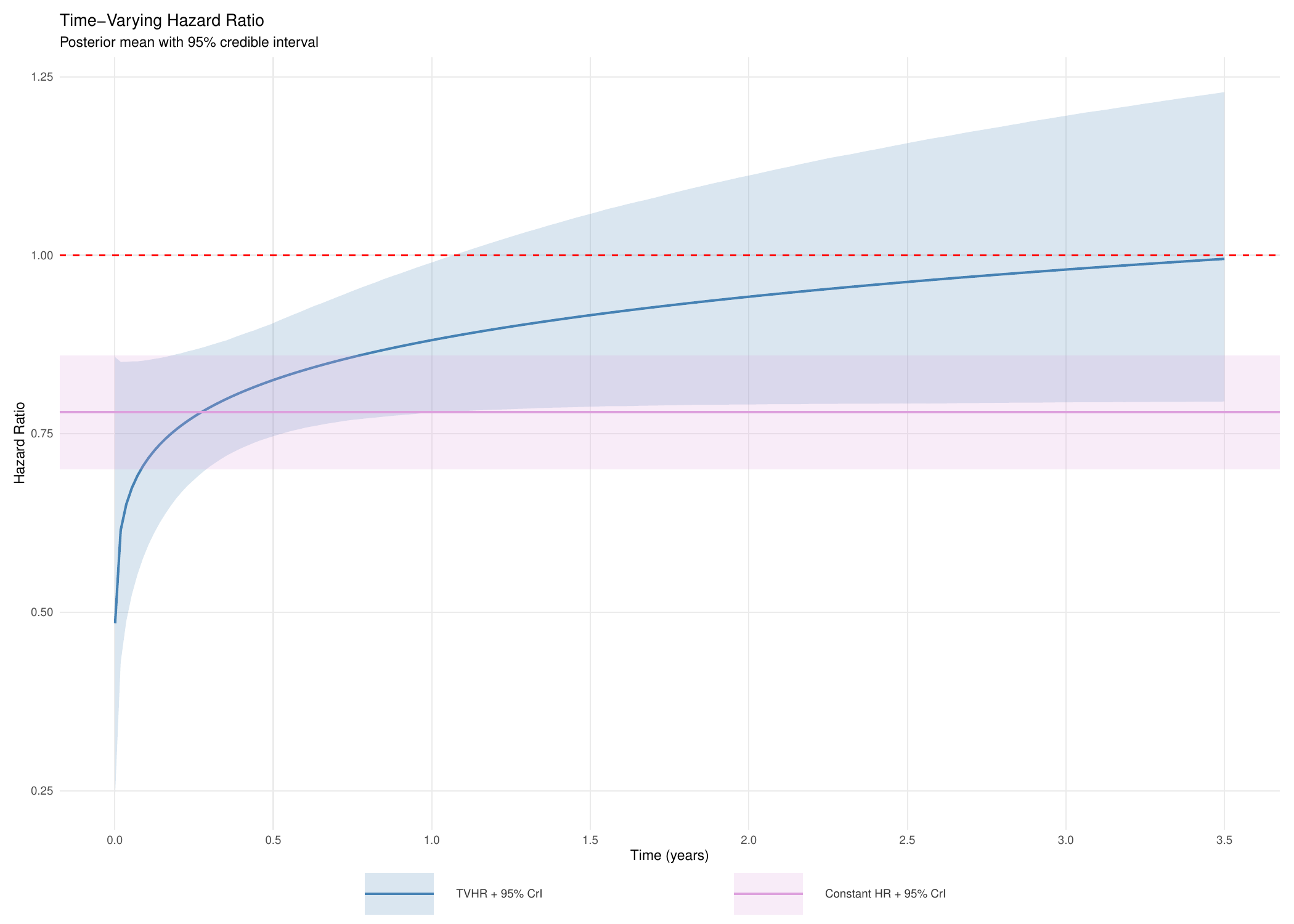}
  \caption{Gastric cancer meta-analysis - pooled time-varying hazard ratio (TVHR) and constant hazard ratio, and associated 95\% credible intervals (CrI).}
  \label{fig:gastricpooledtvhr}
\end{figure}

Figure~\ref{fig:gastricpooledtvhr} and Table~\ref{tab:gastricHR} show the results of fitting the Bayesian 2-stage TVHR bivariate meta-analysis model (\ref{eq:pma3})-(\ref{eq:pma5}) in comparison to a Bayesian 2-stage constant HR meta-analysis model (\ref{eq:pma1}) \& (\ref{eq:pma2}) in terms of the pooled HRs and 95\% CrIs. All values of $\hat{R} <$ 1.01, and ESS $>$1000 per parameter, suggesting that the MCMC chains have converged to a common stationary distribution with reliable posterior inference. The 2-stage TVHR bivariate model produces a posterior mean for the treatment-log(time) interaction parameter of 0.09 with 95\% CrI (-0.005 to 0.193) and a posterior probability that the interaction effect is positive of 0.97, providing supportive evidence that allowing for a TVHR is warranted. It can be seen that, as with the study-specific results (Figure~\ref{fig:gastricstudytvhr}), there are substantial discrepancies between the two approaches at both earlier and later time points. In particular, the upper limit of the 95\% CrI for the TVHR approach is above 1 (no treatment difference) at time points beyond 1 year and by 3 years the posterior mean for the HR is also very close to 1 indicating no benefit of chemotherapy over SoC, whilst the standard 2-stage, approach assuming a constant HR, confers a 22\% relative risk reduction (of progression or death) to Chemotherapy over SoC at {\em all} time points with a 95\% CrI from a 14\% reduction to a 30\% reduction. The sensitivity analysis in Table~\ref{tab:gastricHRsens} showed that the results are invariant to specification of a Half-Normal(0,0.5) prior distribution for $\tau_2$.     

\begin{table}[hbtp]
\centering
\begin{tabular}{lcc} \hline 
Time (Years) & HR & 95\% CrI \\ \hline 
Constant & 0.78 & 0.70 to 0.86 \\
0.5      & 0.83 & 0.75 to 0.91 \\
1.0      & 0.88 & 0.78 to 0.99 \\
1.5      & 0.82 & 0.79 to 1.06 \\
2.0     & 0.94 & 0.79 to 1.11 \\ 
2.5     & 0.96 & 0.79 to 1.15 \\ 
3.0     & 0.98 & 0.79 to 1.19 \\ 
3.5     & 0.99 & 0.79 to 1.23 \\  \hline
\end{tabular}
\vspace{8pt}
\caption{Gastric cancer meta-analysis - pooled constant \& time-varying hazard ratios (HR) and associated 95\% credible intervals (CrI) at specific time points for chemotherapy compared to standard of care estimated using MCMC.}
\label{tab:gastricHR}
\end{table}

%%%%%%%%%%%%%%%%%%%%%%%%%%%%%%%%%%%%%%%%%%%%%%%%%%%%%%%%
% NMA RESULTS
%%%%%%%%%%%%%%%%%%%%%%%%%%%%%%%%%%%%%%%%%%%%%%%%%%%%%%%%

\subsection{Network meta-analysis - Advanced BRAF-mutated Melanoma}

\begin{figure}[htbp]
  \centering
  \includegraphics[width=0.8\textwidth]{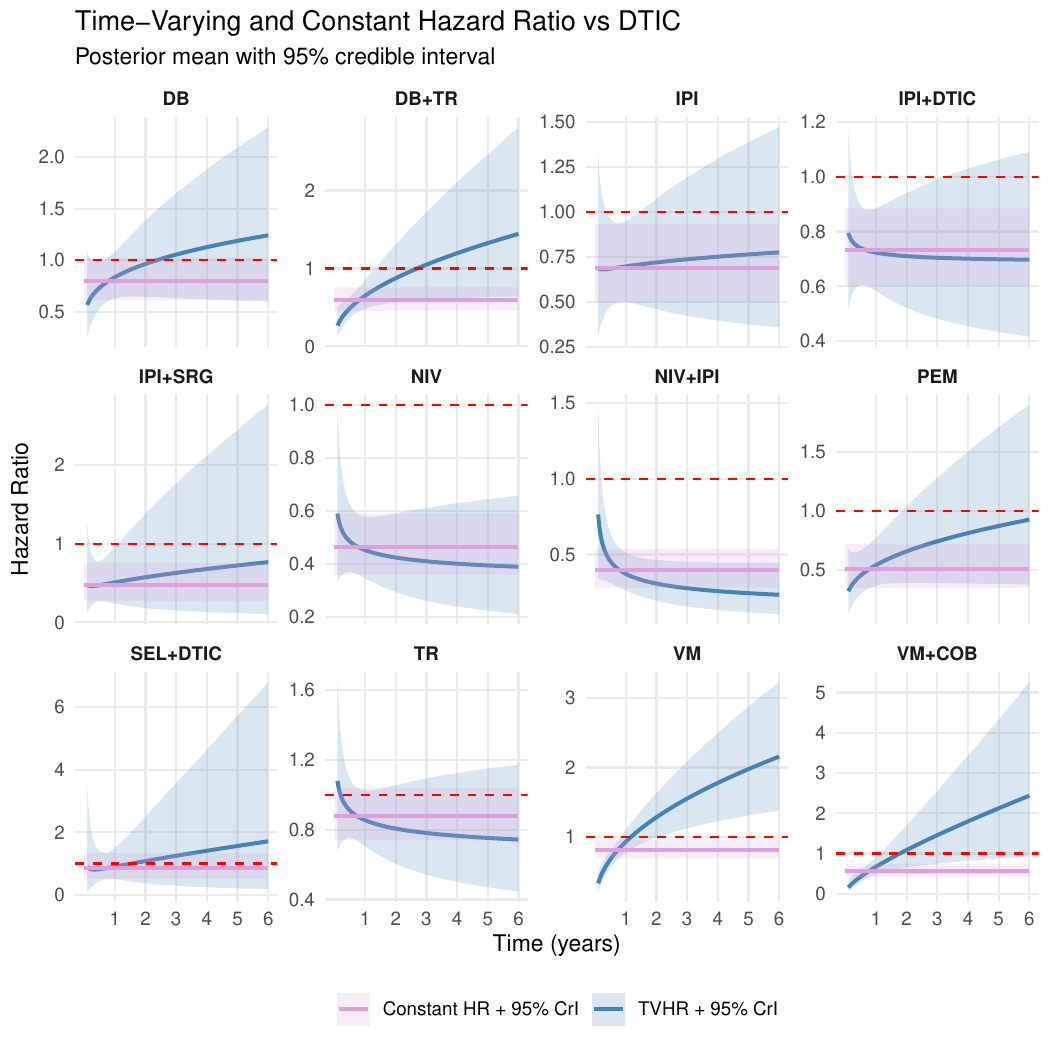}
  \caption{Melanoma network meta-analysis assuming time-varying and constant hazard ratios (HR), and associated 95\% credible intervals (CrI).}
  \label{fig:TVHRNMA}
\end{figure}

Treatment comparisons were informed by single studies, with the exception of NIV + IPI versus IPI, which was informed by 2 studies. As a result, we present findings from fixed effect models as the primary network meta-analysis, which also allow comparison with alternative approaches to address time varying treatment effects as reported in \citep{freeman2022}. Pooled HRs and corresponding 95\% credible intervals (CrI) obtained from the proportional hazards Cox models and TVHR Cox models are illustrated in Figure \ref{fig:TVHRNMA}. Table \ref{tab:melanomaHR} presents the pooled HRs and mean treatment rankings from the network meta-analysis for each treatment compared to dacarbazine (DTIC), both under a constant hazard assumption and at one, two, and five years under a time-varying assumption. Of the 13 studies included, 9 (69\%) were extrapolated beyond the observed period for an average extrapolation period of 2 years (SD: 1.34). Four studies (BREAK-3, CheckMate 067, Keynote 006, and Robert 2011) had an observed period of 5 of more years.  

Under the constant HR assumption, the combination of nivolumab plus ipilimumab (NIV+IPI) demonstrated the greatest efficacy in terms of overall survival, yielding a pooled HR of 0.39 (95\% CrI: 0.28 to 0.54) and the highest mean rank of 1.42 (95\% CrI: 1 to 3). Nivolumab monotherapy (NIV) and pembrolizumab (PEM) also showed substantial benefit, with pooled HRs of 0.47 (95\% CrI: 0.36 to 0.59) and 0.51 (95\% CrI: 0.35 to 0.72), respectively. Several other treatments including ipilimumab (IPI), ipilimumab plus dacarbazine (IPI+DTIC), and vemurafenib plus cobimetinib (VM+COB), demonstrated statistically credible reductions in the hazard of death relative to DTIC, as evidenced by 95\% credible intervals excluding unity.

\begin{figure}[htbp]
  \centering
  \includegraphics[width=0.8\textwidth]{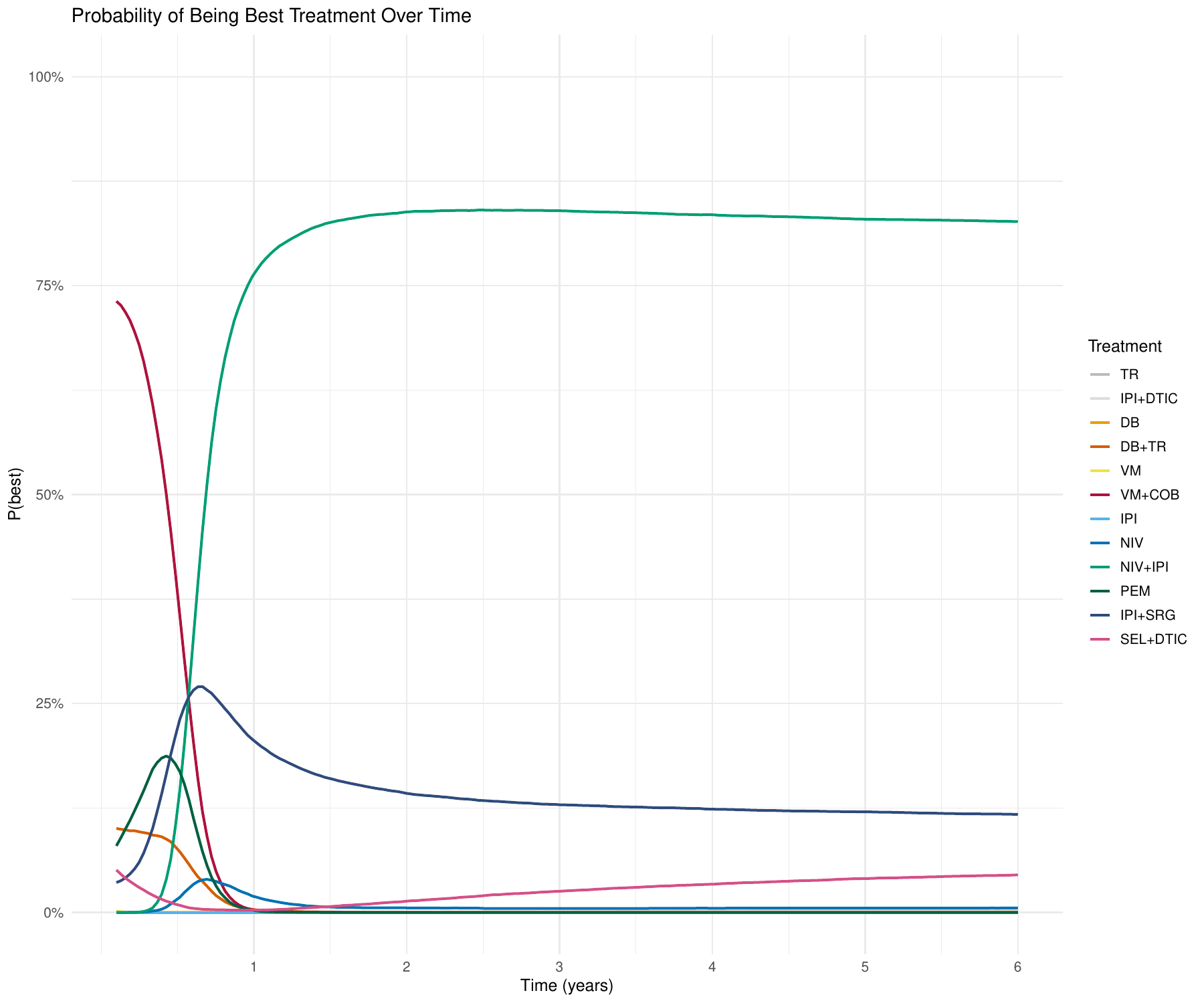}
  \caption{Melanoma network meta-analysis - probability best over time using a TVHR NMA model.}
  \label{fig:pbest_NMA}
\end{figure}

The time-varying analyses revealed important heterogeneity in treatment effects over time. Figures \ref{fig:pbest_NMA} and \ref{fig:heatmap} illustrates how the probability that each treatment was the best overall changes over time. However, it should be noted that rank probabilities can be affected by the number of studies per comparison and the overall network geometry \citep{kibret2014bias}. NIV+IPI maintained consistent superiority from month 7 onwards, with an improving HR from 0.46 (95\% CrI: 0.31 to 0.65) at 6 months to 0.24 (95\% CrI: 0.12 to 0.45) at five years, and retaining the highest mean rank from month 7 (Table \ref{tab:melanomaHR}, Figure \ref{fig:heatmap}). NIV demonstrated a similarly durable and improving effect, with the pooled HR falling to 0.40 (95\% CrI: 0.23 to 0.65) at five years. IPI-based regimens (IPI, IPI+DTIC, IPI+SRG) showed broadly stable pooled HRs over time, suggesting a sustained but non-improving treatment effect (Figure \ref{fig:TVHRNMA}).

In contrast, several targeted therapy combinations exhibited waning treatment effects over time (Figure \ref{fig:TVHRNMA}). VM+COB showed a pronounced pattern, with the highest probability of being the best treatment at  earlier time points (Figures \ref{fig:pbest_NMA} and \ref{fig:heatmap}) but with rising pooled HRs from 0.56 (95\% CrI: 0.41 to 0.74) at the constant time point to 2.13 (95\% CrI: 0.88 to 4.38) at five years. VM alone followed a similar trajectory, with the pooled HR increasing to 1.98 (95\% CrI: 1.30 to 2.89) at five years, suggesting that VM alone may become inferior relative to DTIC over longer time horizons. However, this estimate is based on a log-linear extrapolation far beyond the observed data (2.5 years) and should be interpreted with caution. Dabrafenib based treatments (DB+TR, DB) also demonstrated waning effects (Figure \ref{fig:TVHRNMA}), for example DB+TR had an estimated pooled HR at 6 months of 0.48 (95\% CrI: 0.36 to 0.64), and crossing unity by five years (1.32, 95\% CrI: 0.62 to 2.49), with its mean rank deteriorating from 4.19 (95\% CrI: 1 to 7) to 9.55 (95\% CrI: 4 to 13) (Table \ref{tab:melanomaHR}). Credible interval widths generally increased at longer time points across all treatments, reflecting greater uncertainty in long-term projections, and this was particularly pronounced for treatments with fewer long-term data, such as IPI+SRG and SEL+DTIC (Table \ref{tab:melanomaHR}). All values of $\hat{R} <$ 1.01, and ESS $>$1000 per parameter, suggesting that the MCMC chains have converged to a common stationary distribution with reliable posterior inference. The results were robust to assuming a random effects model (Figures \ref{fig:TVHRNMA_RE}, \ref{fig:pbest_NMA_RE} and \ref{fig:heatmap_RE}). As a result of the limited data to inform estimation of the variability between studies, there is considerable uncertainty in the resulting estimated HRs (Figure \ref{fig:TVHRNMA_RE}).

%\begin{figure}[htbp]
%  \centering
%  \includegraphics[width=0.8\textwidth]{PH_NMA.pdf}
%  \caption{Melanoma network meta-analysis assuming proportional hazards}
%  \label{fig:PHNMA}
%\end{figure}

\begin{table}[hbtp]
\centering 
\begin{tabular}{llcccc} \hline 
Treatment & Time (years) & HR & 95\% CrI & Mean rank & 95\% CrI \\ \hline 
TR        & Constant     & 0.88 & 0.74 to 1.04 & 10.98 & 8 to 13\\
        & 0.5     & 0.91 & 0.75 to 1.10 & 11.58 & 9 to 13\\
          & 1            & 0.86 & 0.71 to 1.03 & 9.93 & 7 to 13\\
          & 2            & 0.81 & 0.60 to 1.06 & 7.22 & 3 to 11\\
          & 5            & 0.75 & 0.47 to 1.15 & 5.79 & 3 to 10\\ 
IPI+DTIC  & Constant     & 0.73 & 0.60 to 0.89 & 8.32 & 5 to 11\\
        & 0.5     & 0.74 & 0.60 to 0.90 & 9.41 & 7 to 12\\
          & 1            & 0.72 & 0.58 to 0.88 & 7.44 & 4 to 11\\
          & 2            & 0.71 & 0.52 to 0.94 & 5.82 & 3 to 10\\
          & 5            & 0.70 & 0.44 to 1.07 & 5.28 & 2 to 9\\ 
DB        & Constant     & 0.80 & 0.61 to 1.03 & 9.54 & 6 to 13\\
        & 0.5     & 0.73 & 0.52 to 0.98 & 9.10 & 5 to 12\\
          & 1            & 0.83 & 0.63 to 1.08 & 9.51 & 6 to 13\\
          & 2            & 0.96 & 0.65 to 1.38 & 9.29 & 5 to 13\\
          & 5            & 1.19 & 0.61 to 2.10 & 8.84 & 4 to 12\\ 
DB+TR     & Constant     & 0.59 & 0.46 to 0.76 & 5.54 & 2 to 8\\
        & 0.5     & 0.48 & 0.36 to 0.64 & 4.19 & 1 to 7\\
          & 1            & 0.65 & 0.48 to 0.85 & 5.83 & 3 to 9\\
          & 2            & 0.87 & 0.56 to 1.30 & 7.90 & 3 to 12\\
          & 5            & 1.32 & 0.62 to 2.49 & 9.55 & 4 to 13\\ 
VM        & Constant     & 0.81 & 0.69 to 0.95 & 9.84 & 7 to 12\\
        & 0.5     & 0.66 & 0.55 to 0.80 & 8.10 & 5 to 11\\
          & 1            & 0.92 & 0.77 to 1.09 & 11.09 & 9 to 13\\
          & 2            & 1.28 & 0.99 to 1.63 & 12.32 & 10 to 13\\
          & 5            & 1.98 & 1.30 to 2.89 & 12.00 & 10 to 13\\ 
VM+COB    & Constant     & 0.56 & 0.41 to 0.74 & 4.87 & 1 to 8\\
        & 0.5     & 0.41 & 0.28 to 0.59 & 2.64 & 1 to 7\\
          & 1            & 0.67 & 0.48 to 0.90 & 6.28 & 3 to 10\\
          & 2            & 1.09 & 0.66 to 1.70 & 10.24 & 5 to 13\\
          & 5            & 2.13 & 0.88 to 4.38 & 11.69 & 7 to 13\\ 
IPI       & Constant     & 0.69 & 0.50 to 0.94 & 7.71 & 5 to 12\\
        & 0.5     & 0.68 & 0.47 to 0.96 & 8.56 & 6 to 12\\
          & 1            & 0.70 & 0.50 to 0.95 & 7.13 & 4 to 11\\
          & 2            & 0.72 & 0.46 to 1.07 & 6.25 & 4 to 11\\
          & 5            & 0.76 & 0.38 to 1.39 & 5.83 & 3 to 10\\ 
NIV       & Constant     & 0.47 & 0.36 to 0.59 & 3.05 & 1 to 5\\
        & 0.5     & 0.49 & 0.37 to 0.63 & 4.49 & 2 to 7\\
          & 1            & 0.45 & 0.35 to 0.58 & 2.61 & 2 to 4\\
          & 2            & 0.43 & 0.30 to 0.59 & 2.44 & 2 to 4\\
          & 5            & 0.40 & 0.23 to 0.65 & 2.54 & 2 to 4\\ 
NIV+IPI   & Constant     & 0.39 & 0.28 to 0.54 & 1.42 & 1 to 3\\
        & 0.5     & 0.46 & 0.31 to 0.65 & 3.55 & 1 to 7\\
          & 1            & 0.37 & 0.26 to 0.51 & 1.26 & 1 to 2\\
          & 2            & 0.31 & 0.19 to 0.46 & 1.17 & 1 to 2\\
          & 5            & 0.24 & 0.12 to 0.45 & 1.18 & 1 to 2\\ 
PEM       & Constant     & 0.51 & 0.35 to 0.72 & 4.02 & 2 to 7\\
        & 0.5     & 0.45 & 0.29 to 0.66 & 3.22 & 1 to 7\\
          & 1            & 0.54 & 0.36 to 0.77 & 4.08 & 2 to 8\\
          & 2            & 0.66 & 0.39 to 1.04 & 5.11 & 3 to 10\\
          & 5            & 0.87 & 0.38 to 1.73 & 6.86 & 3 to 12\\ 
IPI+SRG   & Constant     & 0.47 & 0.28 to 0.76 & 3.21 & 1 to 8\\
        & 0.5     & 0.47 & 0.27 to 0.78 & 3.88 & 1 to 9\\
          & 1            & 0.51 & 0.24 to 0.95 & 3.68 & 1 to 11\\
          & 2            & 0.57 & 0.18 to 1.38 & 4.25 & 1 to 13\\
          & 5            & 0.72 & 0.12 to 2.43 & 4.89 & 1 to 13\\ 
SEL+DTIC  & Constant     & 0.85 & 0.51 to 1.35 & 9.83 & 4 to 13\\
        & 0.5     & 0.82 & 0.43 to 1.41 & 9.68 & 3 to 13\\
          & 1            & 0.90 & 0.51 to 1.50 & 9.77 & 3 to 13\\
          & 2            & 1.08 & 0.36 to 2.50 & 8.86 & 2 to 13\\
          & 5            & 1.55 & 0.21 to 5.73 & 8.21 & 1 to 13\\ 
          \hline\end{tabular}
          \vspace{8pt}
\caption{Melanoma network meta-analysis - pooled constant \& time-varying hazard ratios (HR), mean rank and associated 95\% credible intervals (CrI) at landmark time points for each treatment compared to Dacarbazine (DTIC).}
\label{tab:melanomaHR}
\end{table}

%%%%%%%%%%%%%%%%%%%%%%%%%%%%%%%%%%%%%%%%%%%%%%%%%%%%%%%%
% DISCUSSION 
%%%%%%%%%%%%%%%%%%%%%%%%%%%%%%%%%%%%%%%%%%%%%%%%%%%%%%%%

\section{Discussion}\label{discussion}
The TVHR meta-analysis approach presented here, which assumes a log-linear effect with time, offers a pragmatic and accessible alternative to more complex methods such as fractional polynomial models \citep{jansen2011} for relaxing the proportional hazards assumption. While retaining relative simplicity, this framework can readily be incorporated into cost-effectiveness models (by allowing the TVHR to be applied to the baseline hazard in a target population), making it well-suited to HTA contexts. Where more complex hazard relationships are anticipated, the approach can be extended to incorporate splines on the interaction between  treatment with log-time, providing additional flexibility without sacrificing the core structure of the model.

In practice, the trials contributing to a pairwise or network meta-analysis are often relatively short in duration, and complex hazard functions are not always observed within the follow-up period. Under these circumstances, a log-linear effect with time is likely to be adequate to capture the nature of any time-varying treatment effects. Where individual patient-level covariate information is available, the framework could be further extended by adopting a Multi-Level Network Meta-Regression (ML-NMR) approach \citep{phillippo2020}, enabling more nuanced exploration of treatment effect heterogeneity. 

A two-stage approach has been employed throughout. This is a practical necessity in many applied settings since covariate information is frequently unavailable, particularly where Kaplan-Meier curves have been digitised \citep{guyot2012}, and thus the additional benefits of a one-stage approach are limited in these circumstances. 

From a reporting and validation perspective, a TVHR approach aligns naturally with how results are commonly communicated in the clinical literature, where trials typically report HRs rather than RMST. This makes the approach particularly useful for checking the face validity of pairwise or network meta-analysis results intended to inform decision-making, as outputs remain interpretable to both a clinical and HTA audiences. 

The TVHR framework offers a principled basis for extrapolating treatment effects beyond the observed follow-up period, with appropriate propagation of uncertainty. This is of particular relevance in HTA, where long-term survival extrapolation is routinely required \citep{latimer2022extrapolation}. The importance of capturing time-varying treatment effects is underscored by the emergence of novel immunotherapy combinations in oncology, such as nivolumab plus ipilimumab (NIV+IPI), which exhibit a decreasing HR over time, and one that has profound implications for long-term survival extrapolation and cost-effectiveness conclusions. As such technologies become more prevalent, methods that can appropriately characterise and extrapolate evolving treatment effects are becoming increasingly important for robust decision-making. Critically, the question of treatment waning is frequently raised in HTA submissions, often with arbitrary scenario analyses applied to explore the impact of changing treatment effects over time \citep{trigg2024treatment}. The TVHR approach provides a data-driven basis for exploring such scenarios, grounding long-term projections in the observed trajectory of treatment effects and offering a more transparent and defensible method for characterising uncertainty around treatment waning assumptions and the implications that this might have for guidance.

Whilst the treatment-log(time) interaction approach offers a pragmatic and accessible framework for capturing time-varying treatment effects, several limitations should be acknowledged. First, the assumption of a log-linear relationship between the HR and log(time), whilst often reasonable in the context of relatively short-term RCTs, may not adequately capture more complex or non-monotonic hazard trajectories in all settings, for example in Real World Evidence (RWE) with longer follow-up \cite{handorf2024}. Although extensions incorporating spline terms on the interaction with log-time can address this, such models require greater data to support them and may be difficult to justify in sparse networks.

Secondly, the two-stage approach adopted here, whilst practical, introduces some efficiency loss relative to a one-stage model, particularly where there is substantial between-trial heterogeneity \citep{burke2017meta}. This limitation is compounded in settings where IPD are unavailable and Kaplan-Meier curves have been digitised, as this process introduces additional uncertainty that is not always fully propagated through subsequent analyses \citep{guyot2012, rogula2022method}.

Thirdly, whilst the framework allows for extrapolation of time-varying treatment effects beyond the observed follow-up period, such projections remain inherently uncertain, and the plausibility of long-term extrapolations, including (constrained) spline-based models \citep{jennings2025spline}, should always be assessed using external data and/or in consultation with patient and clinical experts \citep{latimer2013survival}.

Whilst the TVHR model assumes a time-varying HR for each study, in practice some studies may have a HR that is close to being constant, which will be propagated through the model through a treatment-log(time) interaction term close to 0. This is demonstrated for the IPI+DTIC treatment effect, which is informed by a single study (Table \ref{tab:melanomaHR}), where the HRs obtained from the TVHR approach are similar to the constant HR model. In situations where studies report only the HRs and confidence intervals (i.e., no Kaplan-Meier plots, which can be digitised), a TVHR model can not be estimated but they could nevertheless be included by extending the TVHR model to have a hierarchical structure. For example, the study-specific treatment-log(time) parameter could be assumed to be exchangeable with treatments of the same class \citep{owen2015network}.

Several alternative approaches exist for handling non-proportional hazards in the context of meta-analysis and network meta-analysis. Fractional polynomial models \citep{jansen2011} offer a flexible parametric framework for capturing complex hazard functions and have been widely used in HTA \citep{leahy2019pmu109}. However, they can be difficult to implement within network meta-analysis \citep{petersohn2023challenges}, particularly when heterogeneity in the shape of the hazard function exists across trials, and the resulting models can be challenging to incorporate into economic decision models in a transparent and intuitive manner. By contrast, the treatment-log(time) interaction approach assumes a simpler log-linear relationship with log(time), which, whilst less flexible, is more transparent and more readily communicated to a non-specialist audience.

Restricted mean survival time (RMST) based approaches offer a useful alternative that avoids the proportional hazards assumption entirely, and provide an absolute and easily interpretable summary measure of treatment benefit \citep{wei2015meta}. However, trials typically report HRs rather than RMST estimates, and RMST-based network meta-analysis results can be less straightforward to incorporate into cost-effectiveness models or to validate against other published trial results. The TVHR approach, by retaining the HR as the effect measure, aligns more naturally with how evidence is reported, applied (to a target baseline population), and appraised in HTA.

Piecewise exponential models represent another option, allowing the baseline hazard and treatment effects to vary across pre-specified time intervals \citep{crowther2012individual}. Whilst flexible, the choice of cut-points can be arbitrary and results sensitive to their specification \citep{cooney2023extending}. The log-linear interaction approach developed here avoids this by imposing a smooth, continuous relationship between the treatment effect and time, which is both more parsimonious and more amenable to extrapolation.

Bayesian parametric IPD network meta-analysis approaches, including Weibull, Gompertz, log-normal, log-logistic, gamma, and generalised gamma, including multivariate parameterisation of the shape and scale parameters, are alternative approaches \citep{campbell2025}. These approaches capture the time-varying nature directly through the choice of parametric distribution. In HTA, particularly for cost-effectiveness modelling, a common approach is to fit a parametric survival model to a reference treatment arm and apply HRs to obtain survival curves for comparator treatments, or jointly estimate the baseline hazard and treatment effects within a single parametric distributional family \citep{latimer2011nice}. The former lends itself to a TVHR approach, whereas the latter may prefer a parametric network meta-analysis approach. However, the parametric network meta-analysis approach constrains the survival curves to belong to the same parametric family, which may not be the most appropriate distribution across all treatments included in the network. The TVHR approach separates estimation of the baseline survival models and the synthesis of relative treatment effects, which offers a more straight-forward but flexible implementation in economic models.

A flexible network meta-analysis model for time-to-event outcomes using an M-spline approach on the baseline hazard has been proposed \citep{phillippo2025}, with a weighted random walk prior distribution that provides shrinkage, designed specifically to handle complex hazard functions. The M-spline approach is a highly flexible Bayesian one-step model that places a smooth non-parametric prior distribution on the baseline hazard, allowing it to capture arbitrary departures from proportionality, including delayed treatment effects, crossing survival curves, or non-monotone HRs, without specifying a functional form for the time-varying HR. Our approach, by contrast, assumes a specific parametric structure, that is a linear relationship between the log-HR and log(time) within a Cox model, yielding a time-varying HR that varies as a function of log(time). This imposes a more restrictive but more interpretable form. However, as discussed above, to capture more complex hazard shapes, where required, the linear log-time interaction of the TVHR approach can be generalised to a spline function in log(time), and the bivariate synthesis extended to higher dimensions.

%%%%%%%%%%%%%%%%%%%%%%%%%%%%%%%%%%%%%%%%%%%%%%%%%%%%%%%%
% CONCLUSION 
%%%%%%%%%%%%%%%%%%%%%%%%%%%%%%%%%%%%%%%%%%%%%%%%%%%%%%%%

\section{Conclusion}\label{conclusion}
In conclusion, incorporating a treatment-log(time) interaction within a meta-analysis or network meta-analysis of time-to-event outcomes provides a simple and intuitive approach to relaxing the proportional hazards assumption where it is violated in one or more included studies. The resulting estimates can be readily incorporated into economic decision models, supporting more credible long-term extrapolation and cost-effectiveness analyses. The framework is also readily extendable; further developments include a fully Bayesian one-stage implementation, incorporating flexible splines to model treatment-log(time), as well as models accommodating a mixture of IPD and pseudo-IPD within the same analysis.

%%%%%%%%%%%%%%%%%%%%%%%%%%%%%%%%%%%%%%%%%%%%%%%%%%%%%%%%
% DETAILS 
%%%%%%%%%%%%%%%%%%%%%%%%%%%%%%%%%%%%%%%%%%%%%%%%%%%%%%%%

\section*{Acknowledgments}
The authors thank collaborators of the GASTRIC Group and contributors to the melanoma IPD datasets.

\section*{Funding}
RKO is supported by a Health and Care Research Wales (HCRW) - NIHR Advanced Fellowship (HCRW NIHR FS(A)-2023b-RO) and HCRW Senior Research Leaders Award (SRL-25-019).

\section*{Conflict of Interest}

RKO is a member of the National Institute for Health and Care Excellence (NICE) Technology Appraisal Committee, member of the NICE Decision Support Unit (DSU), associate member of the NICE Technical Support Unit (TSU), and a Health and Care Research Wales (HCRW) Senior Research Leader [SRL-25-019]. She has served as a paid consultant to the pharmaceutical industry and international reimbursement agencies, providing unrelated methodological advice. She reports teaching fees from the Association of British Pharmaceutical Industry (ABPI) and the University of Bristol.

KRA is a member of the National Institute for Health and Care Excellence (NICE) Diagnostics Advisory Committee, the NICE Decision and Technical Support Units, and is a National Institute for Health Research (NIHR) Senior Investigator Emeritus [NF-SI-0512-10159]. He has served as a paid consultant, providing unrelated methodological and strategic advice, to the pharmaceutical and life sciences industry generally, as well as to DHSC/NICE, and has received unrelated research funding from Association of the British Pharmaceutical Industry (ABPI), European Federation of Pharmaceutical Industries \& Associations (EFPIA), Pfizer, Sanofi and Swiss Precision Diagnostics/Clearblue. He has also received course fees from ABPI and the University of Bristol, and is a Chief Statistical \& HTA Adviser for Visible Analytics Limited. 

\section*{Data Availability}
All data used are available from the cited publications and available in the {\tt NMA-TVHR} GitHub repository, \url{https://github.com/r-k-owen/NMA-TVHR}.

\bibliographystyle{unsrt}
\bibliography{tvhr}

\appendix

% Relabel figures with S prefix

\renewcommand{\thefigure}{S\arabic{figure}}
\setcounter{figure}{0}  % Reset counter
\renewcommand{\thetable}{S\arabic{table}}
\setcounter{table}{0}
\renewcommand{\thesubsection}{S\arabic{subsection}}
\setcounter{subsection}{0}

\section*{Supplementary Materials}

\subsection{Pairwise meta-analysis - Advanced Gastric Cancer}
\label{SM_PMA}

\begin{figure}[H]
  \centering
  \includegraphics[width=0.8\textwidth]{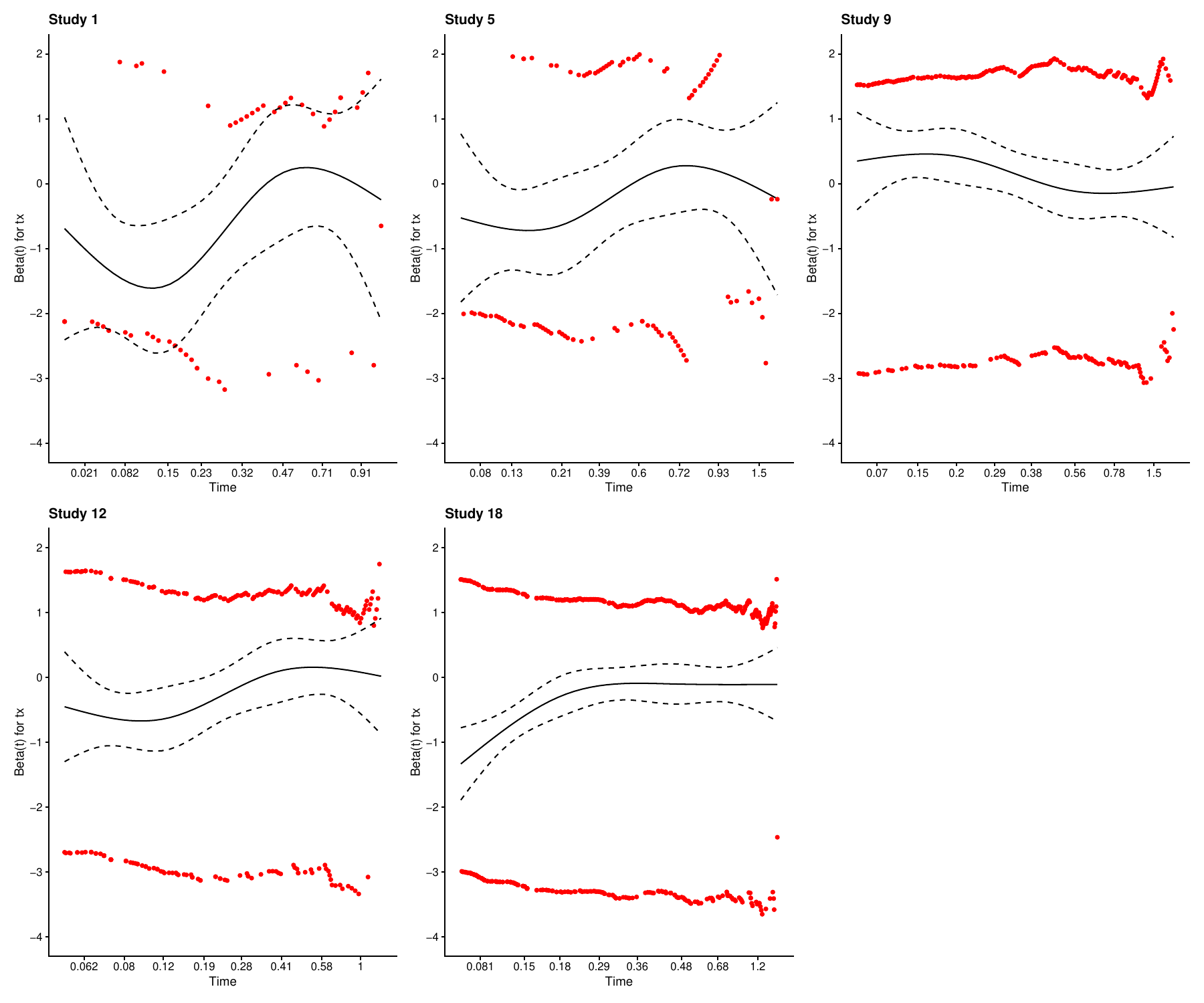}
  \caption{Gastric cancer meta-analysis - Schoenfeld Residual Plots for studies with non-proportional hazards.}
  \label{fig:gastricnonPH}
\end{figure}

\begin{figure}[H]
  \centering
   \includegraphics[width=0.8\textwidth]{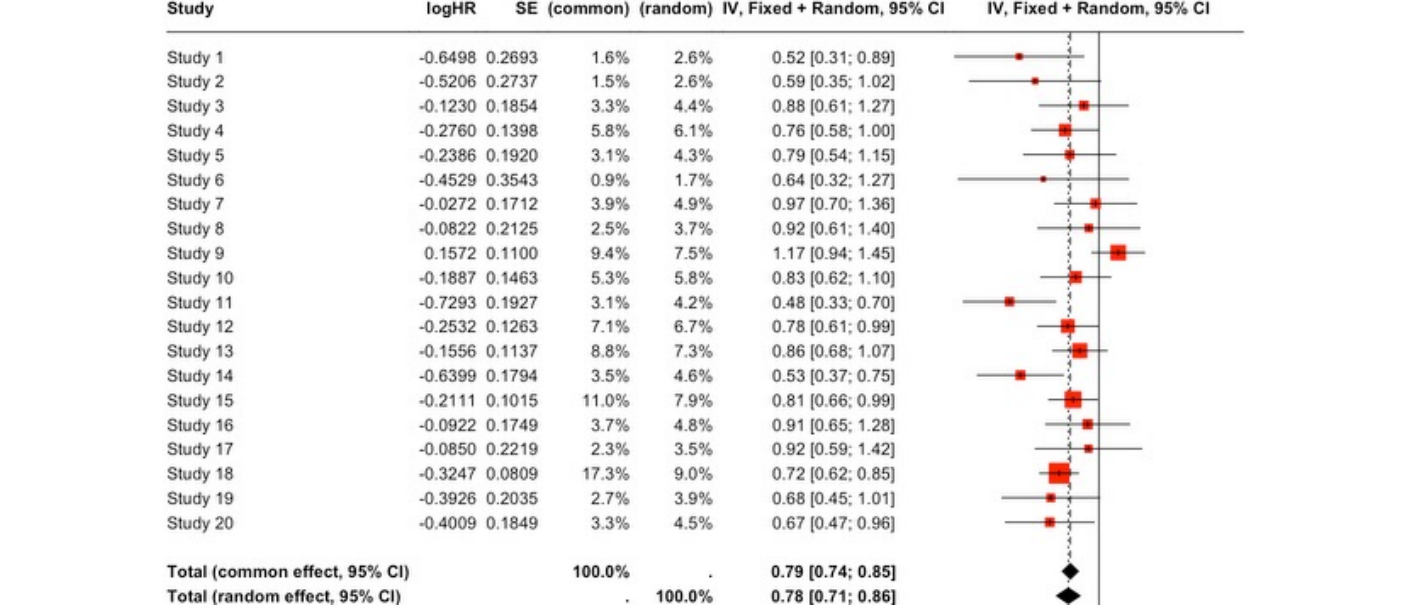}
  \caption{Gastric cancer meta-analysis - two-stage meta-analysis of Hazard Ratios (HRs).}
  \label{fig:gastricma}
\end{figure}

Figure~\ref{fig:gastricma} displays the results of undertaking a frequentist 2-stage meta-analysis (using REML in the {\tt meta} package in {\tt R}) of the log-hazard ratios (and associated standard errors) obtained by applying (\ref{eq:cox}) to each study. The overall pooled HR is 0.78 with 95\% CI (0.71 to 0.86) with statistically significant evidence against an assumption of homogeneity (P=0.01). 

A 1-stage frequentist analysis of the IPD from all 20 studies, and in which each trial is assumed to have a study-specific constant HR and an independent baseline together with a random treatment effect (implemented using {\tt coxme} in {\tt R} \citep{therneau2024coxme,burke2017meta}),  yielded a pooled HR of 0.83 with 95\% CI (0.74 to 0.92).  

\begin{table}[hbtp]
\centering
\begin{tabular}{lcc} \hline 
Time (Years) & HR & 95\% CrI \\ \hline 
0.5 & 0.82 & 0.75 to 0.90 \\
1.0 & 0.88 & 0.78 to 0.99 \\
1.5 & 0.92 & 0.78 to 1.06 \\
2.0 &  0.94 & 0.79 to 1.11 \\
2.5 & 0.96 &  0.79 to 1.16 \\
3.0 &  0.98 & 0.79 to 1.20 \\
3.5 & 0.99 & 0.79 to 1.23 \\ \hline
\end{tabular}
\vspace{8pt}
\caption{Gastric cancer meta-analysis - time-varying hazard ratios (HR) and associated 95\% credible intervals (CrI) at specific time points for chemotherapy compared to standard of care estimated using Half-Normal(0,0.5) prior distribution for $\tau_2$ \& MCMC.}
\label{tab:gastricHRsens}
\end{table}

\subsection{Network meta-analysis - Advanced BRAF-mutated Melanoma}
\label{SM_NMA}

\begin{figure}[H]
  \centering
  \includegraphics[width=0.8\textwidth]{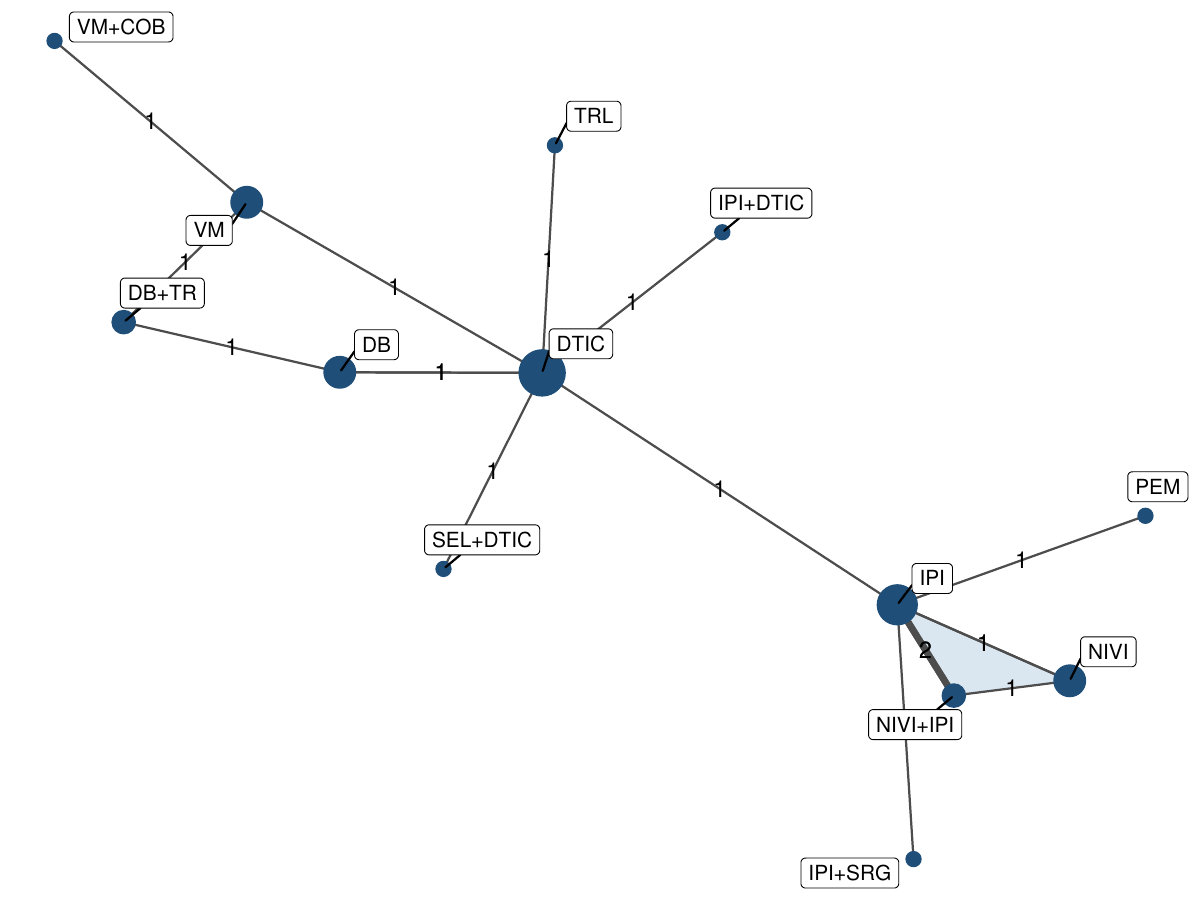}
  \caption{Advanced BRAF-mutated melanoma NMA - Evidence network plot.}
  \label{fig:melanomNMA}
\end{figure}

\begin{figure}[H]
  \centering
  \includegraphics[width=0.8\textwidth]{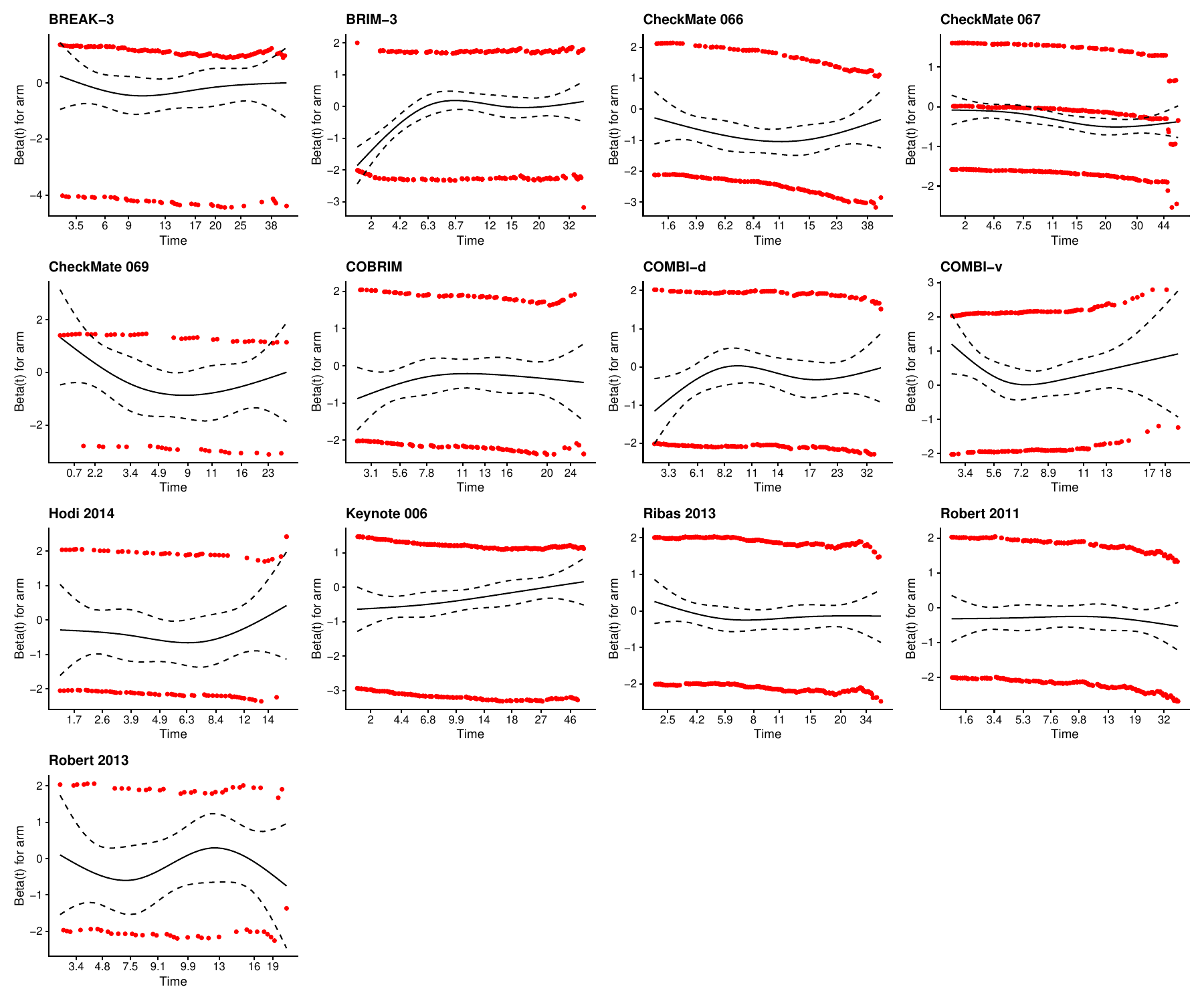}
  \caption{Advanced BRAF-mutated melanoma NMA - Schoenfeld Residual Plots.}
  \label{fig:zph_NMA}
\end{figure}

\begin{figure}[H]
    \centering
    \includegraphics[width=\linewidth]{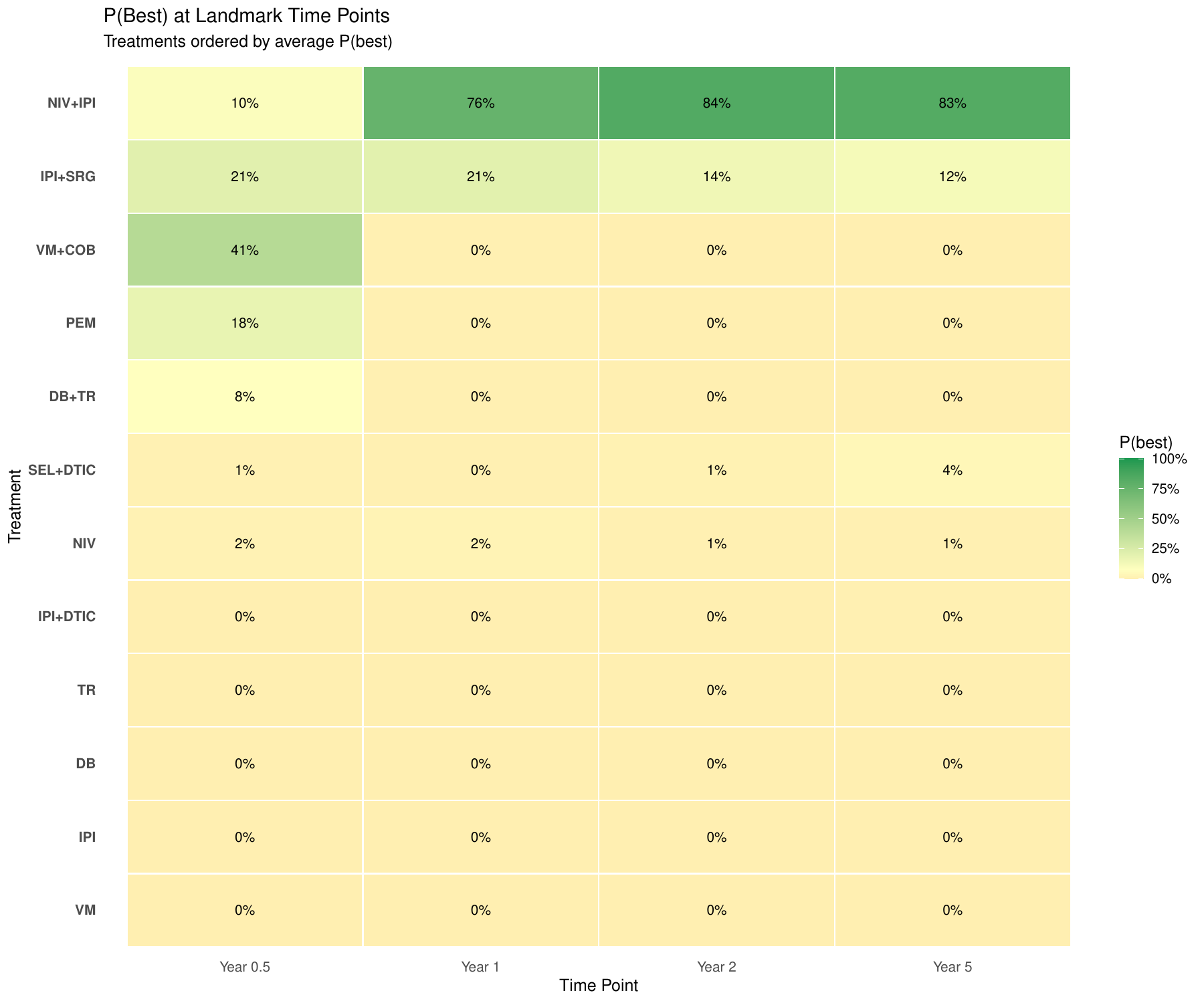}
    \caption{Melanoma network meta-analysis - heatmap of probability best at landmark time points.}
    \label{fig:heatmap}
\end{figure}

\begin{figure}[H]
  \centering
  \includegraphics[width=0.8\textwidth]{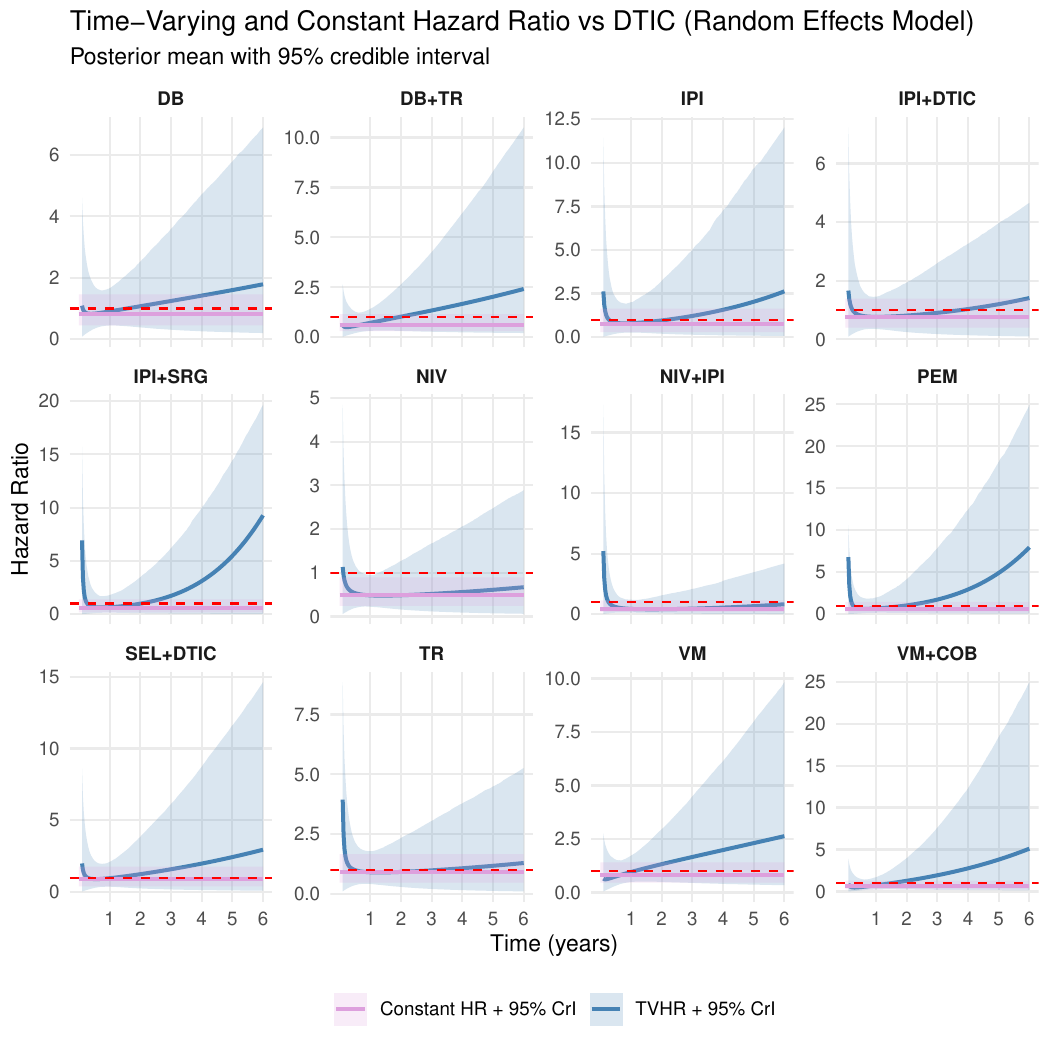}
  \caption{Melanoma network meta-analysis assuming time-varying and constant hazard ratios (HR) assuming random effects, and associated 95\% credible intervals (CrI).}
  \label{fig:TVHRNMA_RE}
\end{figure}

\begin{figure}[H]
    \centering
    \includegraphics[width=\linewidth]{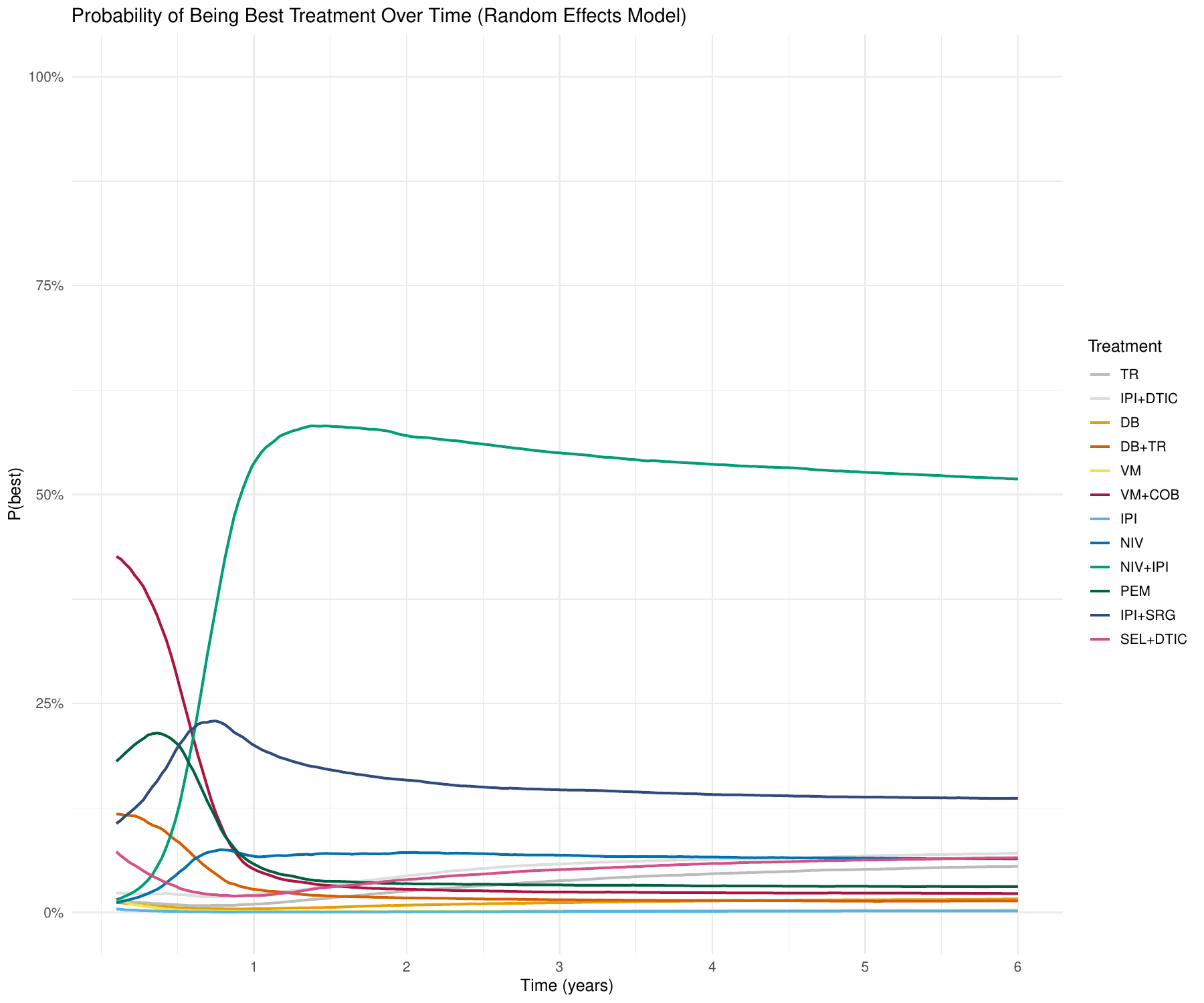}
    \caption{Melanoma network meta-analysis - probability best over time using a TVHR NMA model assuming random effects.}
  \label{fig:pbest_NMA_RE}
\end{figure}

\begin{figure}[H]
    \centering
    \includegraphics[width=\linewidth]{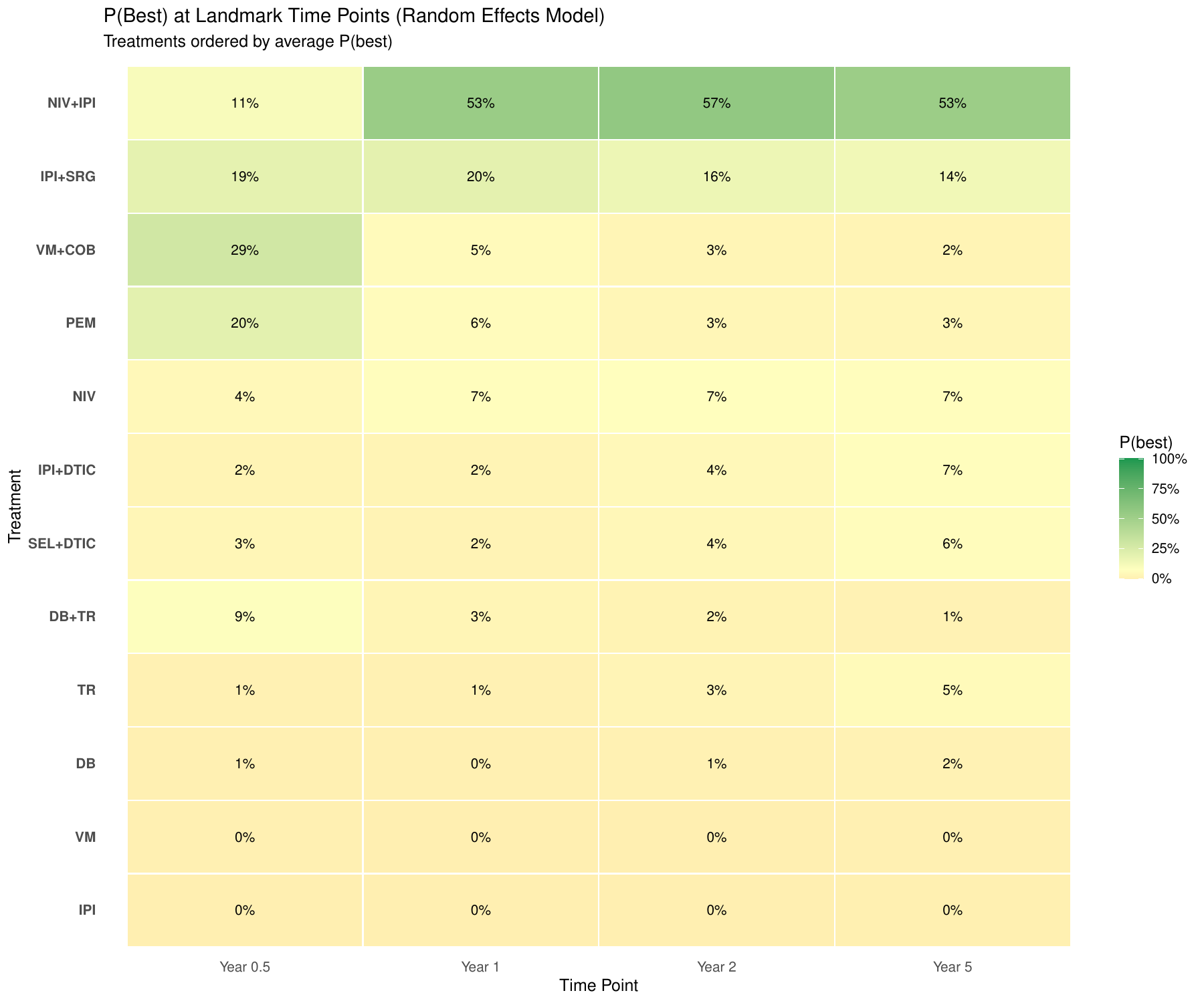}
    \caption{Melanoma network meta-analysis - heatmap of probability best at landmark time points from random effects network meta-analysis.}
    \label{fig:heatmap_RE}
\end{figure}

\end{document}